\documentclass[sigconf,nonacm]{acmart}
\usepackage[T1]{fontenc}
\usepackage{graphicx}
\usepackage{colortbl}
\usepackage[colorinlistoftodos]{todonotes}
\usepackage{array}
\usepackage{makecell}
\usepackage{threeparttable}
\usepackage[ruled,vlined,linesnumbered]{algorithm2e}
\usepackage{setspace}
\usepackage{caption}
\usepackage{subfig}
\usepackage{enumitem}
\usepackage{algorithmicx}
\usepackage{color}
\usepackage{pifont}
\usepackage{xcolor}
\usepackage{array}
\usepackage{svg}
\usepackage{gensymb}
\usepackage{adjustbox}
\usepackage{listings}
\usepackage{xcolor}
\usepackage{algpseudocode}
\usepackage{hyperref}
\usepackage{multirow}
\usepackage{enumitem}
\setlist[itemize]{nosep,leftmargin=*}
\setcopyright{none} 

\graphicspath{{Figures/}}
\DeclareGraphicsExtensions{.pdf,.png,.jpg,.eps} 

\algnewcommand{\IfThenElse}[3]{
  \State \algorithmicif\ #1\ \algorithmicthen\ #2\ \algorithmicelse\ #3}

\begin{document}
\title[]{Not to Break, but to Attest: Adversarial Probes for Privacy-Preserving LLM Verification} 

\author{Cameron Wilding, Mina Shaker, Fatemeh Ganji}
\email{{cwilding, msshaker, fganji}@wpi.edu }
\affiliation{\institution{Worcester Polytechnic Institute}
\city{Worcester, MA}
\country{USA}
}


\begin{abstract}
Post-deployment changes to large language models can alter behavior while leaving routine outputs largely unchanged, creating a challenge for AI governance when model weights are proprietary.
We present a privacy-preserving zk-SNARK-based audit framework that searches for probes designed in the spirit of adversarial examples to amplify logit drift between an approved model and a modified deployment. 
Our framework explores complementary probe families under different access models. 
Token-based probes operate in a black-box setting and require only the input interface, tokenizer, and vocabulary. Embedding-based probes require gray-box access to the embedding interface. 
Stress probes rely on additional interface capabilities but do not require full white-box access to model weights or architecture. 
This range allows probe selection to balance sensitivity, access requirements, and deployment cost.
We evaluate probe constructions across LLM architectures, model-tampering scenarios representative of post-deployment attacks, and GPU platforms. 
Importantly, our experimental results demonstrate that token-based probes consistently deliver the strongest mean sensitivity across models and GPU platforms, although operating in a black-box setting. 
Our Groth16 zk-SNARK workflow remains practical as the probe set scales from 1 to 50, where proving time increases from 1.02 to 1.78 seconds, verification remains near 0.84 seconds, and proof size remains constant. 

\end{abstract}



\maketitle
\thispagestyle{plain}
\pagestyle{plain}

\section{Introduction}

Governance of large language models (LLMs) requires more than evaluating average-case utility or alignment.
It also requires determining whether a deployed model is still the model that was approved, audited, or contractually promised.
This question has become pressing because recent studies and incidents show that hidden modifications can be introduced through model editing, poisoning, backdoor insertion, compromised repositories, or supply-chain attacks on the surrounding software stack.
The concern is not limited to replacing a model.
A more subtle and realistic threat is that a model is modified in a targeted manner to insert a hidden capability, weaken safety behavior in one domain, or induce a narrow behavioral bias, which leave ordinary performance largely unchanged~\cite{li2024badedit, souly2025poisoning, wang2026confguard, souly2025small, ge2025backdoors, dai2025seal, bullwinkel2026detecting}.
Such changes are particularly relevant to governance because they can escape standard evaluation while still altering the model precisely where integrity matters.

A second difficulty is that these hidden changes are typically selective rather than global.
The model may continue to behave normally with routine prompts and public benchmarks, but deviate on carefully chosen inputs, triggers, or dialogue settings.
Prior work repeatedly shows that hidden behavior tends to surface only under particular probing conditions, including token perturbations, embedding-level perturbations, output-confidence scans, explanation prompts, or multi-turn stress tests~\cite{jia2026decoding, tuck2026guided, bullwinkel2026trigger, jamshidi2026adversarial, wang2026confguard, ge2025backdoors}.
These works, however, do not directly provide a governance mechanism for verifying, after deployment and under limited trust, that a model has not drifted away from a committed baseline.
Our work is motivated by this gap.
Specifically, we ask whether synthetic probes can be designed in the spirit of adversarial examples, not to break the model, but to amplify sensitivity to hidden model changes, and thereby support model-integrity verification.

\noindent\textbf{Why probing is the right starting point.}
Prior work addresses individual parts of our setting. LogitLens4LLMs and Tuned Lens use logits and activations to study model state and behavior~\cite{wang2025logitlens4llms,belrose2023eliciting}. Other studies use input perturbations or stress tests to reveal instability, hallucination, unsafe behavior, or poisoning~\cite{jia2026decoding,tuck2026guided,jamshidi2026adversarial,ge2025backdoors}. Model editing, poisoning, and backdoor attacks further show that small, targeted model changes are practical~\cite{bullwinkel2026trigger,li2024badedit,wang2026confguard,souly2025poisoning}. 
Black-box methods such as Model Equality Testing, rank-based uniformity testing, and LLMmap identify or compare models from their outputs~\cite{gao2025model,zhu2026auditing,pasquini2025llmmap}. IMMACULATE, zkGPT, and TeleSparse instead verify all or part of model inference, but require model-specific circuits and greater proving effort~\cite{guo2026immaculate,qu2025zkgpt,maheri2025telesparse}. 

Recent work shows that GPU telemetry can support narrow forms of verification, but not model-level semantic verification.
Telemetry is well suited to observability, profiling, workload characterization, execution-integrity checks, efficiency validation, and post-deployment accountability~\cite{hu2021characterization, chaudhuri2025energon, jain2026minos, weakley2025monitoring, almusaddar2025shadowscope, pedersen2026instant, monfared2026telemetry}.
However, these signals remain indirect with respect to the model itself. 
They can indicate that work ran, that kernels behaved as expected, or that utilization claims are plausible, but they do not establish that a deployed model remains behaviorally consistent with an approved baseline or satisfies a semantic safety specification~\cite{almusaddar2025shadowscope, pedersen2026instant, monfared2026telemetry, intel2026gpuremoteattestation}.
By contrast, probing-based methods have a shorter path to semantic claims because they inspect model-internal or execution-internal state, such as logits, activations, or internal representations~\cite{wang2025logitlens4llms, belrose2023eliciting}.
Our approach follows this latter direction.
Rather than inferring integrity indirectly from hardware behavior, we use synthetic probes designed in the spirit of adversarial examples to amplify logit drift under hidden model changes, and then bind that behavioral signal to a zkSNARK-based verification flow.
Telemetry therefore remains a useful complementary evidence layer, but probing provides the more direct basis for model-level verification.

\noindent\textbf{Private verification of probes.}
The literature also makes clear that plain probe fingerprints are not enough for our setting.
They provide only statistical confidence over the challenged probe set, and their reliability degrades if the provider can learn the probes before the audit challenge, cache answers, or adapt to a fixed public challenge set. 
If one keeps probe verification low-cost, purely functional and statistical, with no cryptographic wrapper, it is inevitable to have weaker secrecy and weaker public verifiability.

A second option is to execute the hidden-probe audit inside a TEE with remote attestation, which is operationally attractive because encrypted probes can be sent into an attested environment, but the resulting assurance is fundamentally about the measured execution environment rather than a mathematical proof of the hidden probe relation~\cite{intel2026sgxdcaporientation, intel2026gpuremoteattestation}.
A third option is multiparty computing-based secure inference, which provides strong privacy during evaluation~\cite{ng2023sok}, but does not naturally yield a small, reusable, non-interactive proof for later third-party verification.
Authenticated data structures such as Merkle trees or polynomial commitment schemes are useful adjuncts for committing to a probe corpus or acceptance metadata, but by themselves they do not prove that the model actually computed the committed function on the hidden probe~\cite{li2006dynamic,kate2010constant}.
This leaves succinct zero-knowledge proof systems, such as the zero-knowledge Succinct Non-interactive Argument of Knowledge (zkSNARK)~\cite{bitansky2012extractable,gentry2011separating,micali2000computationally}, as the most natural cryptographic tool when the objective is to keep probes and outputs hidden while producing portable evidence of consistency.

Among zero-knowledge proof systems, the main trade-off is among verifier efficiency, setup assumptions, flexibility, and proof-system features.
In our setting, the audit relation is small, as a committed model is checked repeatedly on a selected hidden probe batch under a stable acceptance relation.
For this case, a zkSNARK is appealing because it yields a compact, fast, and publicly verifiable proof that reveals neither the probes nor the model responses, and the literature specifically identifies Groth16 as especially attractive when the circuit is stable and succinctness at the verifier is the dominant concern~\cite{groth2016size}. 
Compared with telemetry-only approaches, the resulting claim is semantically stronger because it is tied to the model's functional behavior rather than to indirect resource signals~\cite{cankur2025characterizing, patwari2022dnn}.
Compared with plain probe fingerprints, it preserves probe secrecy and supports third-party verifiability.
Compared with TEE-based alternatives, it offers a cryptographic statement about the hidden probe relation itself rather than a hardware-rooted statement about the measured execution environment~\cite{intel2026sgxdcaporientation, intel2026gpuremoteattestation}.
For repeated, privacy-preserving probe audits, Groth16 is a good fit among the available alternatives~\cite{groth2016size}.

\noindent\textbf{Our approach} differentiates itself from the prior work by exploring the idea of designing probes in the spirit of adversarial examples. 
The goal here is not to induce failure for its own sake.
It is to identify synthetic inputs along which small or targeted model modifications produce disproportionately large logit drift.
This matters because hidden capability insertion, backdoors, safety weakening, and narrow bias are typically selective. 
Usual prompts may reveal them, whereas a carefully designed probe can amplify the effect of the modification and convert a latent change into a measurable signal~\cite{jia2026decoding, tuck2026guided, bullwinkel2026trigger, jamshidi2026adversarial, ge2025backdoors}.

The second idea is to incorporate this probe-derived signal into a zkSNARK-based verification flow, yielding a direct governance benefit. 
At audit time, the verifier issues the probes to the prover, but need not observe the current logits or the proprietary model itself. 
The resulting proof establishes consistency with the committed baseline on the selected probe set without including the current logits or model weights in the public proof transcript. 
In this way, the combination of zkSNARKs and adversarial-style probes supports privacy-preserving model-stability attestation, reduces dependence on semantic prompts or direct weight access, and complements existing approaches such as watermark recovery, backdoor scanning, and alignment stress testing~\cite{dai2025seal, wang2026confguard}. 

\noindent\textbf{Contributions. }
We position our approach against alternatives and argue that it provides a stronger form of model-level evidence under limited trust by making the following contributions. 

\begin{itemize}

    \item We introduce a probe-driven framework for post-deployment AI governance that turns adversarial-style sensitivity from a threat signal into a governance instrument.  
    Such probes characterize sensitivity to modifications that serve as proxies for selective hidden changes, such as hidden capability insertion, backdoor behavior, safety weakening, and narrow behavioral bias.
    \item We develop multiple probe families and compare them in terms of construction, access requirements, deployment cost, and empirical sensitivity. 
    The stress probes are motivated by MLP and attention pathways.  
    Token-based probes require only black-box access plus tokenizer information, embedding-based probes require gray-box access to the embedding interface, and stress-style probes remain hypothesis-driven without requiring full white-box access to weights or architecture. 
    \item We show how to use hidden, high-sensitivity probes rather than semantic prompts or public challenge sets to define a model-focused audit that is harder to game. After selecting the strongest probes, we commit to the approved model’s responses on that hidden probe set and use the resulting behavioral signal as the basis for verification.
    \item We bind this hidden-probe audit to Groth16, yielding a privacy-preserving integrity check.  
    We quantify its practicality through constraint counts, proving and verification time, proof size, scaling with the number of probes, and end-to-end audit latency.
\end{itemize}

\section{Adversary Model and Related Work}
\noindent\textbf{Adversary model. }
We consider a post-deployment audit in which the deployment operator may replace or modify the approved model checkpoint through fine-tuning, adapters, model editing, or direct parameter modification. 
The verifier retains the probe-generation seed, the selected probe set, and the baseline commitment. 
Before the audit challenge, the probes are withheld from the prover. 
At audit time, the verifier regenerates and issues the selected probes to the prover, which therefore learns the challenged probe inputs. 
The proof transcript contains the public commitment, circuit parameters, and Groth16 proof, but does not publicly disclose the probes, raw logits, or private witness values. 
The environment is untrusted with respect to unauthorized model-parameter changes, while the prescribed witness-generation pipeline is assumed to operate faithfully.  
Our focus is on selective behavioral modification rather than generic service failure. 
In particular, we consider adversaries that introduce hidden capability insertion, backdoor or trigger behavior, safety weakening in a narrow domain, or selective behavioral bias, while preserving ordinary benchmark performance and routine user-visible behavior. 

Accordingly, the adversary's goal is to make the served model appear normal on standard prompts and public tests while deviating from the approved baseline on carefully chosen inputs. 
The adversary may realize such changes through targeted fine-tuning, model editing, backdoor insertion, adapter-based modification, or other lightweight post-training changes. 
It may also attempt to exploit the limited observability of remote execution by adapting to known public probes, caching challenge-response pairs, or optimizing against a revealed challenge set. 
Our method is designed to detect this class of hidden behavioral modification under limited trust by using secret probes that amplify sensitivity to such changes and by binding the resulting audit relation to a privacy-preserving proof.

We do not claim full functional equivalence of the deployed model on all inputs. The guarantee is narrower and governance-oriented: the adversary should not be able to convince the verifier that the deployed model is consistent with the committed baseline on the hidden probe set unless that hidden-probe audit relation is actually satisfied.

\subsection{State-of-the-art Probing-based Approaches}
A natural way to organize the probing-based approaches is by the signal being extracted and the purpose for which it is used.
\noindent\textbf{Probing-based representation- and logit-level analysis. }
A first category studies logits and activations as analysis objects.
LogitLens4LLMs~\cite{wang2025logitlens4llms} and Tuned Lens~\cite{belrose2023eliciting} show that meaningful token predictions can be recovered from intermediate layers and that latent prediction trajectories carry information about model behavior.
TaskTracker~\cite{abdelnabi2025get} further demonstrates that activation-level probes can detect behavioral drift during fine-tuning.
These works establish that internal representations and output distributions contain measurable information about model state.
They are, however, primarily interpretability or drift-analysis tools.
They do not construct synthetic governance probes intended to distinguish a clean, deployed model from a stealthily modified one.

\noindent\textbf{Adversarial and stress-style probing.}A second category studies adversarial, perturbation-based, or stress-style probing. 
GPS uses guided masking and embedding instability to detect adversarial text~\cite{tuck2026guided}.
DeP applies semantics-preserving textual perturbations to expose multimodal hallucination through logit and attention drift~\cite{jia2026decoding}.
Adversarial Moral Stress Testing shows that multi-turn adversarial contexts can reveal progressive ethical or safety drift that single-turn tests miss~\cite{jamshidi2026adversarial}.
DeepContext further emphasizes the importance of stateful, multi-turn analysis in adversarial settings~\cite{albrethsen2026deepcontext}.
These studies are close in spirit to ours because they use carefully chosen inputs to amplify latent failures.
Their objectives, however, are robustness, hallucination mitigation, or alignment evaluation.
They do not aim to verify whether a deployed model has been modified in a hidden modification relative to a reference baseline.

\begin{figure}[t]
    \centering
    \includegraphics[width=0.8\columnwidth]{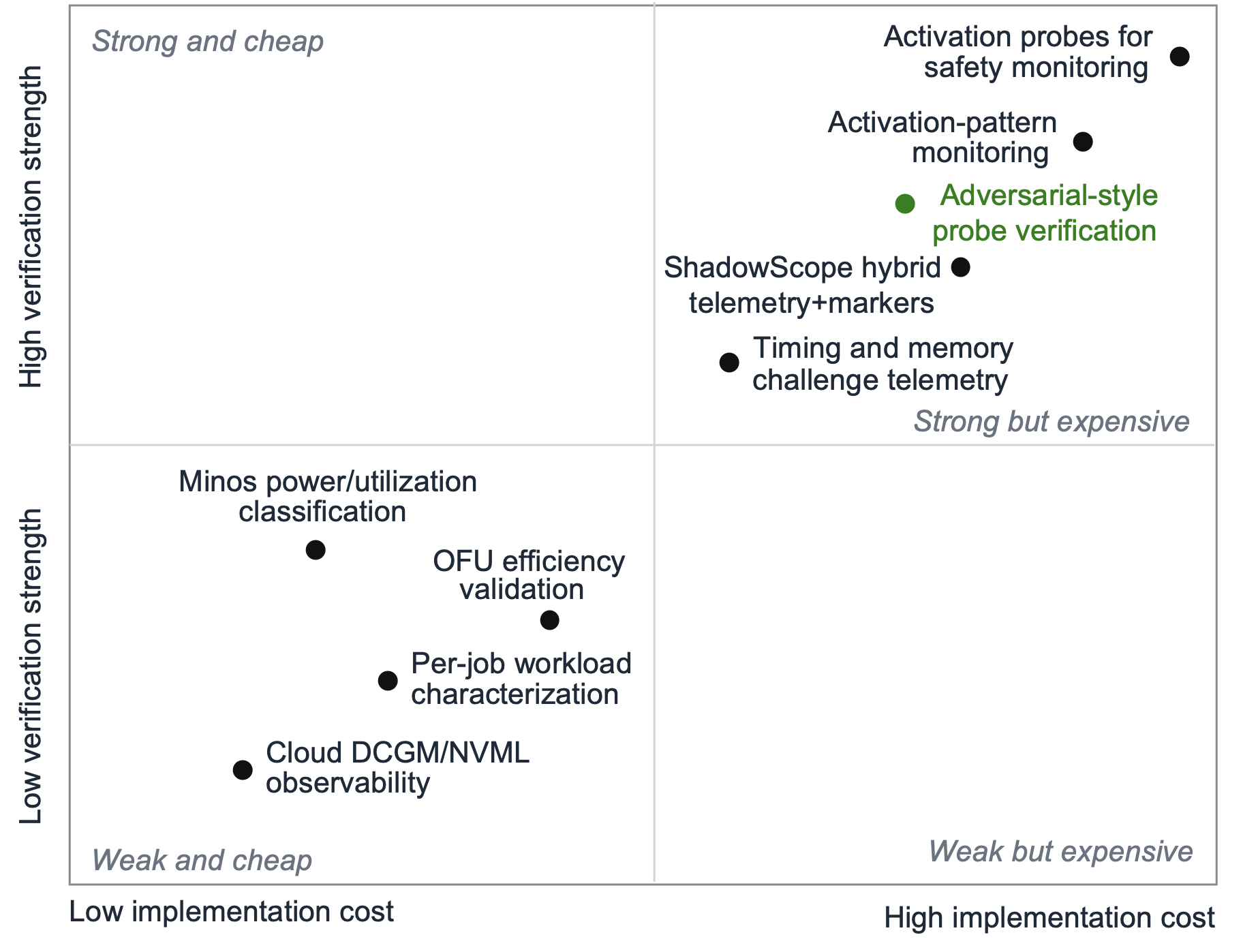}
    \caption{Qualitative comparison of telemetry-based and probing-based approaches along the axes of implementation cost and verification strength. 
    Our adversarial-style probe verification approach is positioned in the strong-but-expensive region. 
    It is more semantically aligned with model-integrity verification than telemetry-only methods, yet less intrusive than full internal activation-monitoring schemes. 
    The placement is qualitative and summarizes the current literature rather than reporting a direct benchmark.}
    \label{fig:cost-vs-verification}
\end{figure}

\noindent\textbf{Probing for backdoor and hidden-modification detection.}
A third category targets hidden malicious modifications directly.
BadEdit shows that small direct weight edits can implant highly effective backdoors with little impact on clean behavior~\cite{li2024badedit}.
ConfGuard detects one specific form of backdoor by identifying unusually stable output-confidence trajectories~\cite{wang2026confguard}.
When Backdoors Speak uses explanation quality, layer-emergence analysis, and attention behavior to distinguish clean from poisoned inputs~\cite{ge2025backdoors}.
Bullwinkel et al.\ recover hidden triggers by scanning the input space and analyzing output-distribution anomalies~\cite{bullwinkel2026trigger}.
Poisoning studies by Souly et al.\ and Anthropic further show that surprisingly small poisoned corpora can suffice across model scales~\cite{souly2025poisoning, souly2025small}.
Together, these results strongly motivate the need for sensitive post-deployment checks.
They nevertheless focus on constructing the attack or on detecting a specific backdoor manifestation, rather than on designing generic synthetic probes for governance-oriented model-integrity testing.

\noindent\textbf{Probing for watermarking and ownership verification.}
A final adjacent category is watermarking and ownership verification. 
SEAL is especially relevant because it explicitly models logit drift under model modification and uses reference queries to estimate the shift in a watermark subspace~\cite{dai2025seal}.
More broadly, watermarking and fingerprinting schemes show that carefully chosen inputs can elicit model-specific responses.
Our setting differs in two respects.
First, we do not assume prior instrumentation, ownership signals, or watermark structure.
Second, we do not seek to prove ownership.
We instead seek to determine whether hidden changes have occurred after a model has been approved, using synthetic probes chosen precisely because they amplify sensitivity to such changes.
To the best of our knowledge, the combination of adversarial-style synthetic probes and zkSNARK-based verification for AI governance has not been studied in this form.

\noindent\textbf{Probing vs. telemetry for verification. }
GPU telemetry has a real but limited connection to verification.
Most telemetry systems and studies are designed for observability, profiling, scheduling, workload characterization, or security auditing, rather than for model-semantic validation~\cite{hu2021characterization, chaudhuri2025energon, jain2026minos, weakley2025monitoring}.
The strongest verification-oriented claims in this literature are substantially narrower.
ShadowScope validates GPU kernel execution against a golden reference using PMU-side-channel traces~\cite{almusaddar2025shadowscope}.
OFU validates framework-reported efficiency and FLOPs-accounting claims using hardware counters~\cite{pedersen2026instant}.
Timing and Memory Telemetry on GPUs for AI Governance uses challenge-response telemetry to infer utilization and support post-deployment accountability~\cite{monfared2026telemetry}.
These are meaningful forms of verification, but they verify execution integrity, accounting correctness, or utilization evidence, not whether a deployed model remains behaviorally consistent with an approved baseline or satisfies a semantic safety specification~\cite{almusaddar2025shadowscope, pedersen2026instant, monfared2026telemetry, intel2026gpuremoteattestation}.
This limitation is structural rather than accidental.
Telemetry observes resource behavior: utilization, memory and HBM use, power, clocks, thermal patterns, interconnect traffic, PMU traces, or challenge-response latency~\cite{nvidia2026dcgm, amd2026amdsmi, oneapi2026levelzero, googlecloud2026dcgmmetrics, microsoft2026aksgpuobservability, alibaba2026clusterdata}.
Such signals are well matched to questions about whether work ran, how intensively a GPU was used, whether a kernel deviated, or whether platform-state evidence is valid~\cite{almusaddar2025shadowscope, pedersen2026instant, monfared2026telemetry, intel2026gpuremoteattestation}.
They are not well matched to questions about whether a model's outputs are semantically correct, robust, fair, safe, or policy-compliant.
By contrast, probing-based methods, e.g.,~\cite{wang2025logitlens4llms, belrose2023eliciting}, inspect model-internal or execution-internal state, including activations, representations, inserted markers, or explicit runtime predicates, and therefore have a shorter path to semantic claims.
This distinction is central to our setting.
Our method does not infer model integrity indirectly from hardware activity.
Instead, it uses synthetic probes designed in the spirit of adversarial examples to amplify logit drift under hidden model changes, and then binds that behavioral signal to a zkSNARK-based verification flow.
In this sense, telemetry remains a useful complementary evidence layer, but probing provides the more direct basis for model-level verification.

\section{Background Information on zkSNARK}
\noindent\textbf{Notation.}
Let $f_\theta(x)$ denote the next-token logit vector produced by model parameters $\theta$ on input $x$, measured at the last token position.
Let $\theta'$ denote the deployed model under audit.
For a candidate probe $x$, we define the instability score as
$\Delta(x)=\|f_\theta(x)-f_{\theta'}(x)\|_2$.
Let $P=\{p_1,\dots,p_K\}$ denote the final hidden probe set of size $K$, where the top-$K$ probes are the $K$ probes with the largest $\Delta(x)$ within the selected probe family.
Running the approved baseline model on $P$ yields $\mathrm{baseline\_logits}$, and running the deployed model on $P$ yields $\mathrm{current\_logits}$.
We define the public commitment as $C=\mathrm{Hash}(\mathrm{baseline\_logits})$ and the recomputed audit value as $C'=\mathrm{Hash}(\mathrm{current\_logits})$.

In the zkSNARK, the public statement $\phi$ contains $C$ and public circuit parameters, while the witness $w$ contains $\mathrm{current\_logits}$ and auxiliary circuit values.
The audit relation $R(\phi,w)$ holds iff the hidden-probe audit circuit is satisfied and enforces $C'=C$.
The verification circuit is represented as an arithmetic circuit, equivalently as an R1CS instance, where each R1CS constraint is one algebraic rule of the form $(\text{linear combination})\times(\text{linear combination})=(\text{linear combination})$.
The witness size denotes the amount of private input data supplied to the prover, reported either as the number of witness field elements or as its serialized size.

\noindent\textbf{Zero-knowledge succinct non-interactive argument of knowledge (zkSNARK).} 
A zkSNARK lets a prover convince a verifier that a statement is true, while revealing nothing beyond that fact and keeping the proof short and the verification cost low.
Groth16 is especially attractive in practice because, for arithmetic-circuit statements, it gives a preprocessing SNARK with a proof consisting of only three group elements and verification based on a single pairing-product check using three pairings in total~\cite{groth2016size}.
In our setting, the statement is not ``the model is correct'' in a semantic sense; rather, it is that the committed audit relation over the selected probe set holds.
Concretely, after the probe set is identified, the prover uses Groth16 to show that the current probe outputs satisfy the same commitment relation as the approved baseline, without including the probes, current logits, or model weights in the public proof transcript. 

\noindent\textbf{Formal definition adapted to our setting.}
Following Groth~\cite{groth2016size}, let $R$ be a binary relation over statements $\phi$ and witnesses $w$.
A publicly verifiable non-interactive zero-knowledge argument of knowledge consists of four probabilistic polynomial-time algorithms $(\mathsf{Setup}, \mathsf{Prove}, \mathsf{Verify}, \mathsf{Sim})$.
They are defined as follows:
$(\sigma,\tau)\leftarrow \mathsf{Setup}(R)$, where $\sigma$ is the common reference string and $\tau$ is the simulation trapdoor;
$\pi\leftarrow \mathsf{Prove}(R,\sigma,\phi,w)$, where $(\phi,w)\in R$ and $\pi$ is the proof;
$b\leftarrow \mathsf{Verify}(R,\sigma,\phi,\pi)$, where $b\in\{0,1\}$ indicates acceptance or rejection; and
$\pi\leftarrow \mathsf{Sim}(R,\tau,\phi)$, where the simulator produces a proof for $\phi$ without using the witness~\cite{groth2016size}.
The scheme satisfies \emph{completeness}, meaning that an honest prover with a valid witness is always accepted; \emph{knowledge soundness}, meaning that any prover that convinces the verifier can be used to extract a valid witness; and \emph{zero knowledge}, meaning that the proof reveals nothing beyond the truth of the statement~\cite{groth2016size}.

In our setting, the statement $\phi$ contains the public audit data, the commitment $C=\mathsf{H}(\mathsf{baseline\_logits})$ with $\mathsf{H}(\cdot)$ denoting the hash function, together with any other public circuit parameters.
The witness $w$ contains the private audit data, namely the current probe outputs, and any auxiliary values needed by the circuit.
The relation $R(\phi,w)$ is satisfied exactly when the witness corresponds to valid probe evaluations whose committed value matches the public baseline commitment.
Thus, the Groth16 proof certifies that the deployed model remains consistent with the committed baseline on the selected high-sensitivity probe set, while keeping the current logits private and excluding the probes and logits from the public proof transcript. 

\noindent\textbf{Arithmetic-circuit view. }
Groth16 is defined for arithmetic-circuit satisfiability, equivalently for quadratic arithmetic programs, where a statement fixes public inputs and a witness fixes the remaining private variables needed to satisfy the circuit constraints~\cite{groth2016size}.
This view matches our use case directly.
Once the top-$K$ probes are determined, the verification task is compiled into an arithmetic circuit that takes the current private probe outputs, recomputes the commitment value inside the circuit, and enforces equality with the public baseline commitment.
The resulting proof is succinct, zero knowledge, and directly tied to the probe-based audit relation rather than to hardware telemetry or semantic prompt evaluation.

\section{Methodology}
\label{sec:Methodology}

\subsection{Problem Setting and Core Idea}

Adversarial examples are typically studied as a threat because they reveal directions in input space along which a model is unusually sensitive.
Our starting point is that this same phenomenon can be turned into something useful.
Instead of using such inputs to break the model, we use synthetic probes designed in the spirit of adversarial examples to expose hidden model changes for governance purposes.
The intuition is that a stealthily modified model may still appear normal under routine evaluation, yet respond differently on carefully chosen inputs that excite sensitive internal directions.
If such probes can be identified in advance, they can serve as compact tests of model integrity.
This is the central idea of our methodology.

The setting we consider is motivated by the type of hidden changes that are most relevant to AI governance.
We are not concerned with arbitrary noise or wholesale replacement of the model alone.
Instead, we focus on targeted and stealthy modifications such as hidden capability insertion, backdoor behavior, safety weakening in a narrow domain, or selective behavioral bias.
These changes are attractive precisely because they can preserve ordinary behavior on standard prompts while altering the model where integrity matters most.
For this reason, average-case evaluation is insufficient.
A useful governance mechanism should instead stress the model along carefully chosen directions where such modifications are more likely to become behaviorally visible.

This observation also explains why our methodology does not rely on semantic prompts.
If the goal is to detect a hidden change without already knowing its content, then the probes should not depend on task-specific meaning or prior knowledge of the inserted behavior.
They should instead depend only on the model interface and on the fact that some inputs reveal sensitivity better than others.
In this regard, our work treats probes as governance instruments rather than as attacks.
A good probe is not one that makes the model fail in an obvious way.
A good probe is one for which a small or targeted model modification induces a disproportionately large and measurable change in the model's logits.

\subsection{Access Assumptions and Relation to Telemetry}

An important consequence of this design is that probe generation requires only limited knowledge of the deployed model.
In the most basic case, token-based probes require only the tokenizer and vocabulary size.
If embedding-space probes are used, the required additional information is limited to whether the interface supports \texttt{inputs\_embeds} and, if so, the embedding dimension.
More generally, the methodology assumes only a narrow contract between the model owner and the deployment platform, namely the inference interface, the admissible input format, and deterministic evaluation settings for collecting logits at a fixed position.
It does not require access to the weights, the training data, or the model's full internal architecture.

This limited-knowledge setting is important for governance because, in practice, a model owner and a GPU provider may be willing to agree on interface-level guarantees without exposing the model internals.
It also distinguishes our setting from telemetry-oriented approaches such as \cite{monfared2026telemetry}.
That line of work is valuable because it asks whether timing-, memory-, and workload-dependent observables can reveal GPU utilization in adversarial environments without trusted firmware or vendor-controlled telemetry.
In particular, \cite{monfared2026telemetry} uses PoW-inspired measurements, VDF-style sequential workloads, GEMM-based tensor-core stress, and VRAM-residency tests to infer hardware engagement.
These measurements provide resource-level visibility, which is important for accountability and post-deployment monitoring.
At the same time, their signal is fundamentally hardware-level rather than model-level.
That is, such telemetry may indicate that a GPU is active, contended, memory-resident, or executing workloads consistent with AI use, but it does not establish whether the deployed model itself remains behaviorally consistent with an approved baseline.
Our methodology is designed to address this latter question directly.
We do not ask whether the hardware appears busy.
We ask whether the model's behavior, under a secret and highly sensitive probe set, remains consistent with a committed baseline.
In this sense, the two directions are complementary.
Telemetry provides evidence of hardware engagement, whereas our probes provide evidence of behavioral consistency with an approved model.

\begin{figure}[t]
    \centering
    \includegraphics[width=\columnwidth]{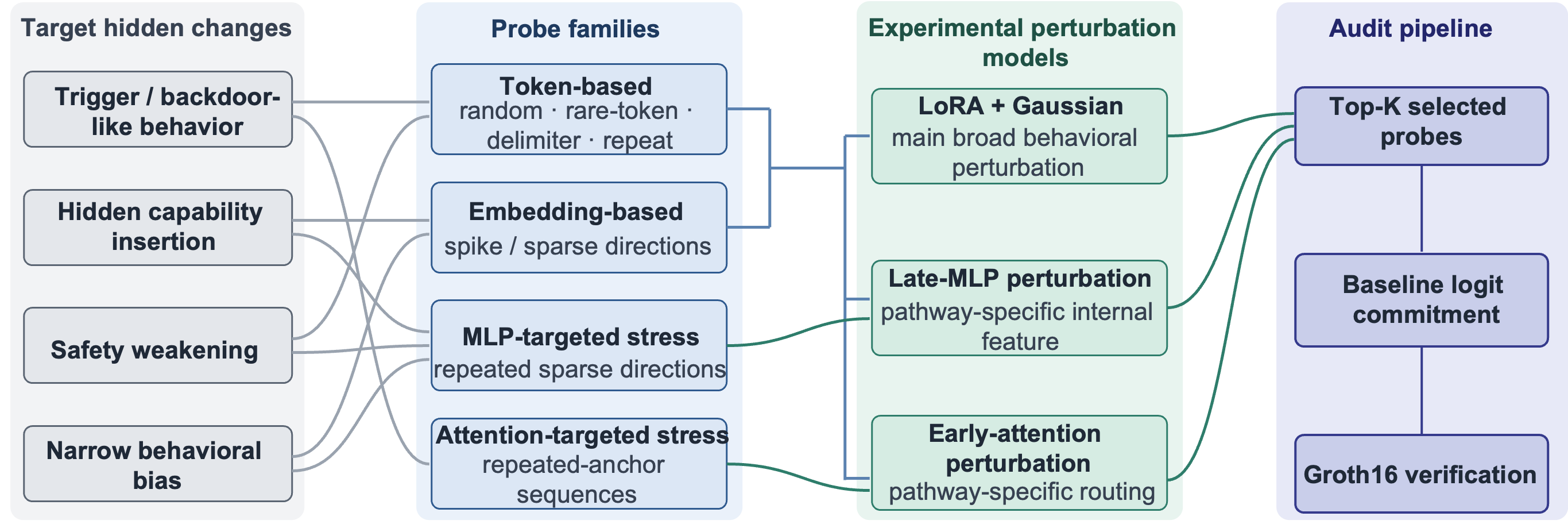}
    \caption{Overview of the proposed probe-driven AI-governance pipeline. Target hidden-change classes motivate the design of four probe families. These families are evaluated under perturbation models. After selecting the top-$K$ probes, the baseline logits are committed via $C=\mathrm{Hash}(\text{baseline logits})$, and Groth16 is used to verify that the deployed model remains behaviorally consistent with the committed baseline without including the probes or logits in the public proof transcript.}
    \label{fig:probing_pipeline}
\end{figure}

\subsection{Probe Families, Access Models, and Design Trade-offs}

Different probe families are necessary because the hidden changes relevant to AI governance are heterogeneous in both their mechanisms and their visibility.
Trigger- or prefix-like modifications, hidden capability insertion, safety weakening, and narrow behavioral bias need not surface through the same inputs, the same internal pathway, or the same level of model access.
We therefore separate probes by both \emph{targeted change class} and \emph{access model}.
Token-based probes target broad functional sensitivity, including trigger-like and general behavioral drift, and are the cheapest to deploy because they require only black-box access together with tokenizer and vocabulary information.
Embedding-based probes target localized, direction-specific representation changes and are more expressive, but they require gray-box access to the embedding interface and embedding dimension.
Stress-style probes are the most hypothesis-driven. 
MLP-targeted probes are motivated by the hypothesis that repeated sparse directions may increase sensitivity to changes in feature-transformation pathways. 
Attention-targeted probes are designed to examine whether repeated anchor patterns may increase sensitivity to routing-, prefix-, and trigger-related changes. 

These stress-style probes do not require full white-box access to weights or architecture, but they are operationally heavier because they assume stronger interface support and a more specific model of where hidden changes are likely to manifest.
This distinction also determines experimental cost: token probes are the simplest and cheapest to generate and evaluate; embedding probes add interface requirements and a larger search space; and stress probes trade additional design complexity for testing pathway-motivated constructions. 
Structuring the probe space in this way is central to our methodology, because it lets us compare families not only by empirical sensitivity, but also by the kind of hidden change they are meant to reveal, the level of model access they require, and the deployment cost they impose. 

\begin{figure}[t]
    \centering
    \includegraphics[width=\columnwidth]{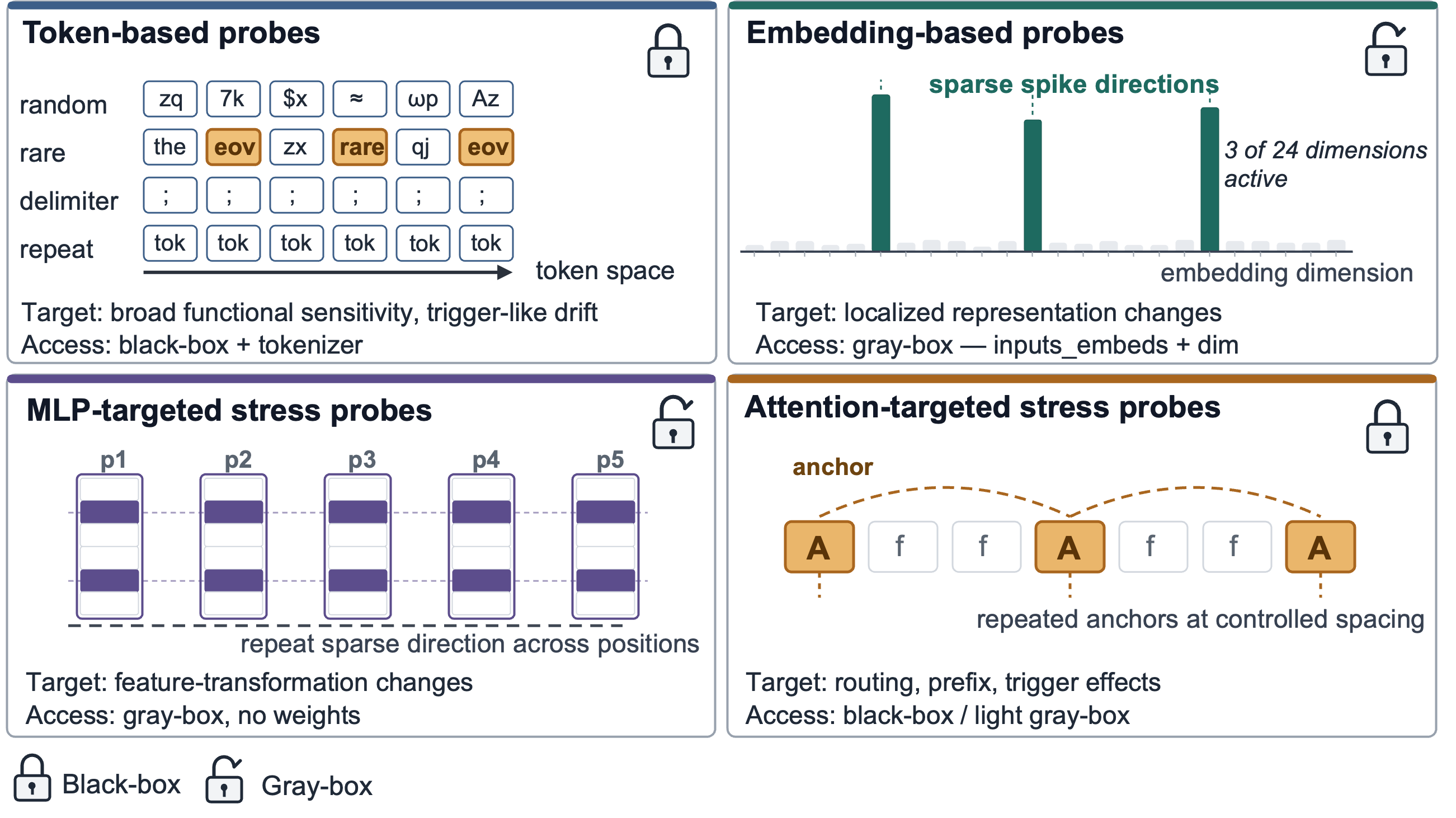}
    \caption{Probe families, constructions, and access requirements. 
    None of the probe families require full white-box access to the weights or the architecture.}
    \label{fig:probe-families-access}
\end{figure}

Formally, let $f_{\theta}(x)$ denote the next-token logit vector produced by model parameters $\theta$ on input $x$, and let $\theta'$ denote a modified version of the same model.
For a candidate probe $x$, we define the instability score
$
\Delta(x)=\|f_{\theta}(x)-f_{\theta'}(x)\|_2.
$
A probe is therefore useful when $\Delta(x)$ is large.
This definition captures the precise role of the probe in our setting.
We do not ask whether the input is meaningful to a user.
We ask whether it exposes a behavioral difference between the original and modified model.
Throughout this work, logits are recorded at a fixed position, namely the next-token logits after the last input token.
This provides a single, consistent measurement point across all probe families, captures the cumulative effect of the entire input after all attention and MLP interactions have propagated through the model, and simplifies the subsequent proof construction.

Based on this design, we organize candidate probes into three families.
The first family operates directly in token space and includes Random token sequences, rare-token patterns, delimiter patterns, repeat patterns, and sequence-length variants.
These probes are useful because they are simple, non-semantic, and easy to generate.
They require only the tokenizer and vocabulary size and provide a natural baseline for whether token-space constructions alone can reveal hidden model changes.
The second family operates in embedding space.
From this family, we retain only spike/sparse embedding probes.
These probes activate a small number of embedding dimensions while leaving the rest inactive.
They are especially aligned with the adversarial-example intuition because they stress a small set of directions that may disproportionately affect model behavior.
The third family contains stress-style probes that intentionally overstimulate specific internal computation pathways.
We retain two subclasses.
The first is MLP-targeted probes, which are designed to stress feature-transformation pathways.
The second is attention-targeted probes, which are designed to stress context-routing behavior by repeated exposure to salient positions. 

Within the token-based family, Random sequences serve as the unstructured synthetic baseline. Rare-token, delimiter, and repeat patterns test whether imposing structure yields additional sensitivity. 
We do not assume that a particular structured construction will dominate across models; candidate generation is therefore followed by model-specific probe search and selection.

\subsection{Perturbation Models and Probe Generation}

The modified models against which these probes are evaluated are chosen to reflect the type of hidden changes we want to study.
This point is important.
Our probes are aimed at targeted, behavior-changing modifications, not arbitrary or extremely weak drift.
For this reason, Gaussian-only perturbations are insufficient as the main modified model.
They represent largely unstructured noise.
Such perturbations are useful as a control, but they do not resemble a stealthy adversary who wants to change one capability or one behavior while leaving the rest of the model largely normal.
A LoRA-only modification is also insufficient in our setting.
Although it is more structured, it is often too selective and too localized to the adapter's own domain. 
In our preliminary experiments, the LoRA-only variant produced only subtle logit drift on the non-semantic probes and therefore did not provide a sufficiently representative test case for comparing probe-family sensitivity to the targeted model changes considered here.
As we discussed earlier, such a change may affect only prompts closely related to the adapter's training purpose while leaving non-semantic probes almost unchanged.
Therefore, for the main experiments, we use the original model as the baseline and the original model with LoRA plus Gaussian perturbation as the primary changed model.
This combined setting preserves a structured targeted change, but makes its effect broad enough for the probes to detect.

For the stress-style probes, we use localized perturbations chosen to test their pathway-specific design hypotheses. 
MLP-targeted probes test whether repeated sparse directions and repeated-anchor constructions provide a family-level advantage when the pathway motivating their design is modified.
MLP-targeted probes are evaluated against late-MLP perturbations.
This reflects the idea that targeted capability changes, safety weakening, or narrow behavioral edits are often expressed through internal feature transformations.
Attention-targeted probes are evaluated against early-attention perturbations.
This reflects the idea that trigger sensitivity, prompt-prefix effects, and context-routing changes are often associated with how the model distributes attention across positions.
This matching between probe construction and perturbation type is part of the methodological rationale.
The purpose is not to maximize diversity of perturbations for its own sake, but to study whether particular probe families are especially sensitive to the hidden changes they were designed to reveal.

Candidate probes are generated in large numbers within each family.
For token-based probes, this means random and structured token sequences with controlled lengths and token patterns.
For spike/sparse embedding probes, this means sparse vectors with varying sparsity levels and amplitudes.
For MLP-targeted probes, sparse directions are repeated across multiple positions to intensify pressure on feature-transformation pathways.
For attention-targeted probes, repeated anchors are inserted at controlled spacings within filler sequences so that the model must repeatedly decide which positions matter.
Each candidate is evaluated on the original model and on the relevant modified model, and $\Delta(x)$ is then computed.
At this stage, the synthetic inputs become adversarial-style probes in the governance-relevant sense.
They are not adversarial because they are malicious.
They are adversarial-style because they are selected for producing unusually large behavioral separation under hidden model changes.

\subsection{Probe Selection and Baseline Commitment}

Probe selection proceeds in two stages.
First, we determine the best family of probes.
This is done by comparing token-based, embedding-based, and stress-style probes against the relevant modified model and identifying which family gives the strongest and most stable signal.
The family-level criteria are the ones we established in our earlier discussion: the strongest family should have the largest mean or median $\Delta(x)$, a clear separation in the corresponding boxplots or histograms, a strong top-$K$ sensitivity curve, and consistency across repeated runs.
Second, once the best family has been identified, we retain the final top-$K$ probes within that family according to $\Delta(x)$. 
The perturbation models are used during calibration as representative modification classes for identifying sensitive probes; they are not assumed to reveal the specific modification that may later occur in deployment. 
Here, $K$ is not an arbitrary parameter.
It governs a real trade-off.
Larger $K$ typically gives a stronger and more robust signal, but it also increases witness size, constraint count, and proving time.
Thus, the methodology already anticipates the later zkSNARK cost study when fixing the final probe set. 
The score \(\Delta(x)\) is used only during off-circuit probe selection; the subsequent Groth16 audit is binary and checks equality of the resulting quantized response commitments rather than the magnitude of logit drift.

Once the final probe set is determined, the trusted baseline model is run on those probes, and the resulting logits are stored as $\text{baseline\_logits}$.
$C=\mathrm{H}(\text{baseline\_logits})$ is a commitment to these logits, the public reference used in later audits.
The probes themselves and the raw logits remain private.
This design choice is essential to the method's governance value.
If the probes were public, an adaptive adversary could attempt to optimize around them.
If the raw logits were public, the audit would reveal more information about the model than is desirable in many practical settings.
The commitment therefore serves as the bridge between behavioral sensitivity and privacy-preserving verification.

\subsection{Private Verification with Groth16}
\begin{figure*}[t]
    \centering
    \includegraphics[width=0.9\textwidth]{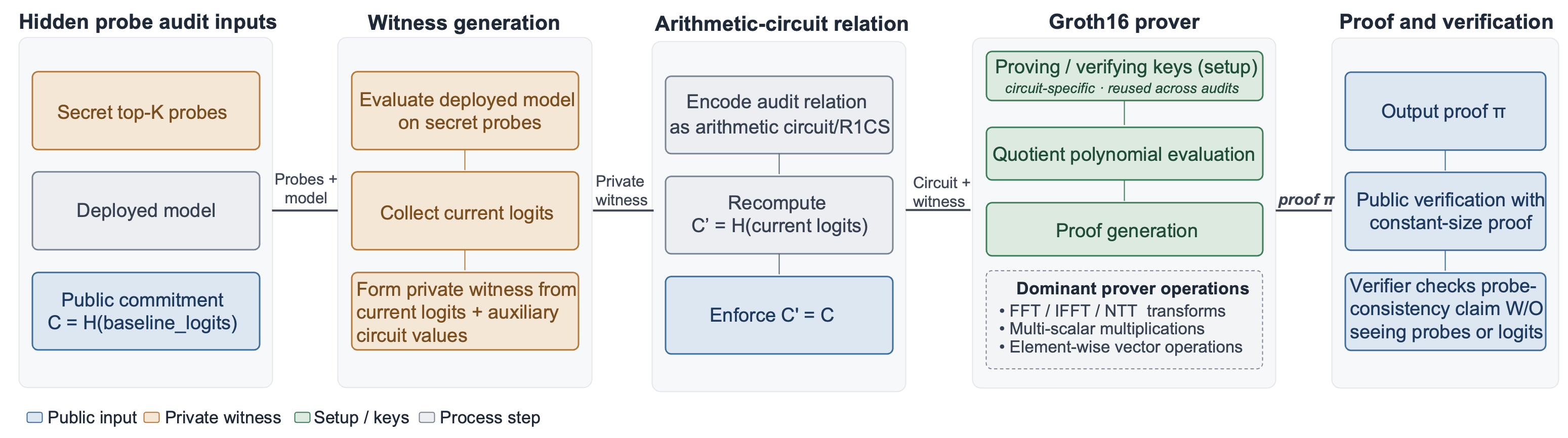}
    \caption{Groth16 in the hidden-probe AI-governance audit pipeline. 
    The audit verifier retains the selected probe set and withholds it from the prover until the audit challenge, at which point the probes are issued for model evaluation. 
    The witness remains private, while the public inputs contain only \(C\) and the circuit parameters. 
    The circuit is fixed once the probe set and acceptance relation are predetermined, allowing the setup to be reused.  
    The public proof transcript does not disclose the probes, raw logits, or private witness values. 
    The audit verifier already knows the probes and verifies consistency of the resulting committed response with the approved baseline.}
    \label{fig:groth16_probe_audit}
\end{figure*}

At audit time, the verifier regenerates the hidden probes from the retained seed and issues them to the prover. 
The prover evaluates the current model, deterministically quantizes and hashes the resulting logits, and uses the resulting \(K\) field elements as the private witness. 
Inside the circuit, the prover computes $C'=\mathrm{Hash}(\text{current\_logits})$ and enforces $C'=C$. 
The verifier thereby checks behavioral consistency with the approved baseline without learning the current logits or model weights. 
The role of zkSNARKs in this methodology is therefore very precise.
They do not help discover the probes.
They do not improve the sensitivity of the probes.
Instead, they allow the governance claim to be verified without including the probes, current logits, or model weights in the public proof transcript. 
In other words, once adversarial-style probes have converted hidden model changes into a measurable behavioral signal, Groth16 turns that signal into a privacy-preserving integrity check. 

In our setting, Groth16 is attractive not only because the verifier sees a constant-size proof and performs only a small pairing check, but also because the prover workflow matches the structure of our hidden-probe audit relation. 
Once the top-$K$ probes and acceptance relation are determined, the resulting circuit is fixed as well. 
The prover then instantiates that circuit with a concrete assignment $\{a_0,a_1,\dots,a_m\}$, where the public portion encodes the audit statement, including the commitment $C=H(\mathrm{baseline\_logits})$, and the witness portion encodes the current private probe outputs together with any auxiliary circuit values. From this assignment, the prover forms the QAP polynomials
$A(X)=\sum_{i=0}^{m} a_i U_i(X)$,
$B(X)=\sum_{i=0}^{m} a_i V_i(X)$, and
$W(X)=\sum_{i=0}^{m} a_i W_i(X)$.
The assignment is valid exactly when the target polynomial $t(X)$ divides $A(X)B(X)-W(X)$, that is, when there exists a quotient polynomial $h(X)$ such that
$A(X)B(X)-W(X)=h(X)t(X)$.

Operationally, proving proceeds in two stages. In the first stage, the prover computes $h(X)$. As in high-performance Groth16 implementations, this is done by working across multiple polynomial representations: the prover begins from the QAP evaluations, applies inverse FFTs to recover coefficient form, evaluates again over a coset domain using FFTs, and then computes the quotient polynomial in evaluation form by pointwise operations of the form
$h(X)=\bigl(A''(X)\circ B''(X)-W''(X)\bigr)/t(\eta\cdot \omega)$
over the chosen domain. In the second stage, the prover forms the proof elements themselves from the witness assignment, the quotient polynomial, and the proving key. Concretely, the proof consists of three group elements,
$\pi=([\pi_A]_1,[\pi_B]_2,[\pi_C]_1)$,
where $[\pi_A]_1$ and $[\pi_B]_2$ are linear combinations of the witness assignment with the proving-key encodings of $U_i(x)$ and $V_i(x)$, together with the standard randomization terms, and $[\pi_C]_1$ combines witness-dependent terms, the quotient contribution through $h(x)$, and the corresponding randomization corrections. In practice, these steps are dominated by a small number of large MSMs, FFT/IFFT/NTT-style polynomial transforms, and element-wise vector operations, exactly as suggested by the implementation-oriented decomposition in Figure~\ref{fig:groth16_probe_audit}. The verifier, by contrast, uses the proof and the public input accumulator to check a single pairing-based relation.

This structure is well aligned with our use case. 
The audit circuit is stable across repeated checks, so the setup artifacts can be reused. 
The witness changes from audit to audit in the current hidden-probe outputs, but the verifier-side cost remains very small even though the prover-side work is substantial. 
For this reason, Groth16 is a particularly good fit for repeated hidden-probe audits, where the same probe-consistency relation is proved many times and succinct verification is more important than universal setup flexibility.

The contrast with \cite{monfared2026telemetry} becomes especially clear at this stage.
Their telemetry framework is designed to reveal utilization-related observables such as contention, memory pressure, and workload alignment at the GPU level.
Our framework is designed to reveal hidden behavioral modifications at the model level.
As a result, their measurements are valuable even when the model is unknown, whereas ours are valuable when the model must remain fixed relative to an approved baseline.
Conversely, our method requires only limited interface-level knowledge of the model, whereas telemetry methods rely on the hardware's architectural behavior.
These are different trust surfaces.
The former is tied to the inference contract.
The latter is tied to the accelerator.
For post-deployment AI governance, this distinction matters because evidence of utilization alone does not imply model integrity, and evidence of model integrity alone does not imply compliant hardware use.

\subsection{Evaluation}
During probe design, the main question is which probe family most strongly exposes hidden changes.
After the final top-$K$ set has been selected, that question is answered. 
The focus then shifts from behavioral sensitivity to cryptographic practicality. 
Accordingly, the Groth16 evaluation uses only the final selected probe set and measures the cost of verifying a realistic governance check.
We report the total number of R1CS constraints for the Poseidon Merkle-tree and root-equality circuit. 
Off-circuit quantization, serialization, and SHA-256 hashing are accounted for as part of witness preparation. 
We further report proving time, verification time, proof size, scalability with the number of probes $K$, and end-to-end audit latency,
$
T_{\mathrm{audit}}=
T_{\mathrm{inference}}+
T_{\mathrm{witness\ prep}}+
T_{\mathrm{prove}}+
T_{\mathrm{verify}}.
$
These measurements are not merely implementation statistics.
They quantify whether the proposed governance mechanism remains practical once the strongest probe family has been identified. 

Overall, the methodology is built around a simple but important shift in perspective.
Adversarial-style sensitivity is usually studied as a weakness of the model.
Here, we use it as a source of leverage.
Synthetic probes are chosen because they amplify the behavioral effect of hidden modifications.
A commitment is then made to the approved model's response on those probes.
Finally, Groth16 is used to verify that the deployed model still matches that committed response without publicly disclosing the probes or the current logits. 
In this way, the methodology combines behavioral sensitivity and cryptographic privacy into a single framework for post-deployment AI governance. 
\section{Results}

\subsection{Implementation and Experimental Setup}
\label{Implem}

We implemented the proposed attestation framework in Python. Model inference was performed on an NVIDIA RTX A4000 GPU via the HuggingFace Transformers library, using the meta-llama/Llama-3.2-1B model in float16 precision. All remaining components — probe generation, logit hashing, commitment computation, witness generation, and proof verification — execute on CPU.

Probe generation follows a two-phase design. In the selection phase, four probe families are generated from a fixed seed and scored against their respective model variants: token probes against a LoRA-finetuned and Gaussian-perturbed variant, sparse embedding probes against the same variant, MLP-targeted probes against a late-layer MLP noise variant, and attention-targeted probes against an early attention noise variant. The family achieving the highest mean $\Delta(x)$ is selected, and the top-$K$ probes within that family are retained. In the attestation phase, only these $K$ probes are used.

The commitment pipeline operates at two levels. At the outer level, each probe's logit vector is quantized to 64-bit integers and compressed to a single BN254 field element via SHA-256, reducing 128,256-dimensional logit vectors to scalar field elements outside the circuit. At the inner level, the $K$ field elements are combined into a single commitment $C$ via a Poseidon~\cite{poseidon} Merkle tree, padded to the next power of two. Both levels use the circomlibjs reference implementation to ensure exact agreement with the circom circuit.
The zk-SNARK layer is implemented using circom 2.2.3 and snarkjs 0.7.6 under the Groth16 proof system. For each value of $K \in \{1, 5, 10, 20, 50\}$, a separate circuit is compiled encoding a Poseidon Merkle tree of the appropriate depth, with the $K$ probe hashes as private witnesses and $C$ as the sole public input. The trusted setup used the Hermez powers-of-tau ceremony at powers 14 and 15, sufficient for circuits up to 32,768 wires. Proving and verification times are reported as the mean and standard deviation over five independent runs, excluding model inference and witness preparation. 
In our implementation, the current logits are prover-private witness-generation data rather than direct R1CS inputs. 
Quantization, serialization, and SHA-256 hashing are performed outside the circuit, and the resulting \(K\) BN254 field elements form the private circuit inputs. 
The circuit reconstructs only their Poseidon Merkle root and enforces equality with the public baseline commitment. 
Accordingly, the reported Groth16 costs correspond to this digest-level commitment circuit and assume, consistently with our adversary model, that the prescribed witness-generation pipeline operates faithfully.

\subsection{Logit Drift Distribution within Families}
\label{subsec:fam1}

\begin{figure}[tb]
  \centering
  \includegraphics[width=\linewidth]{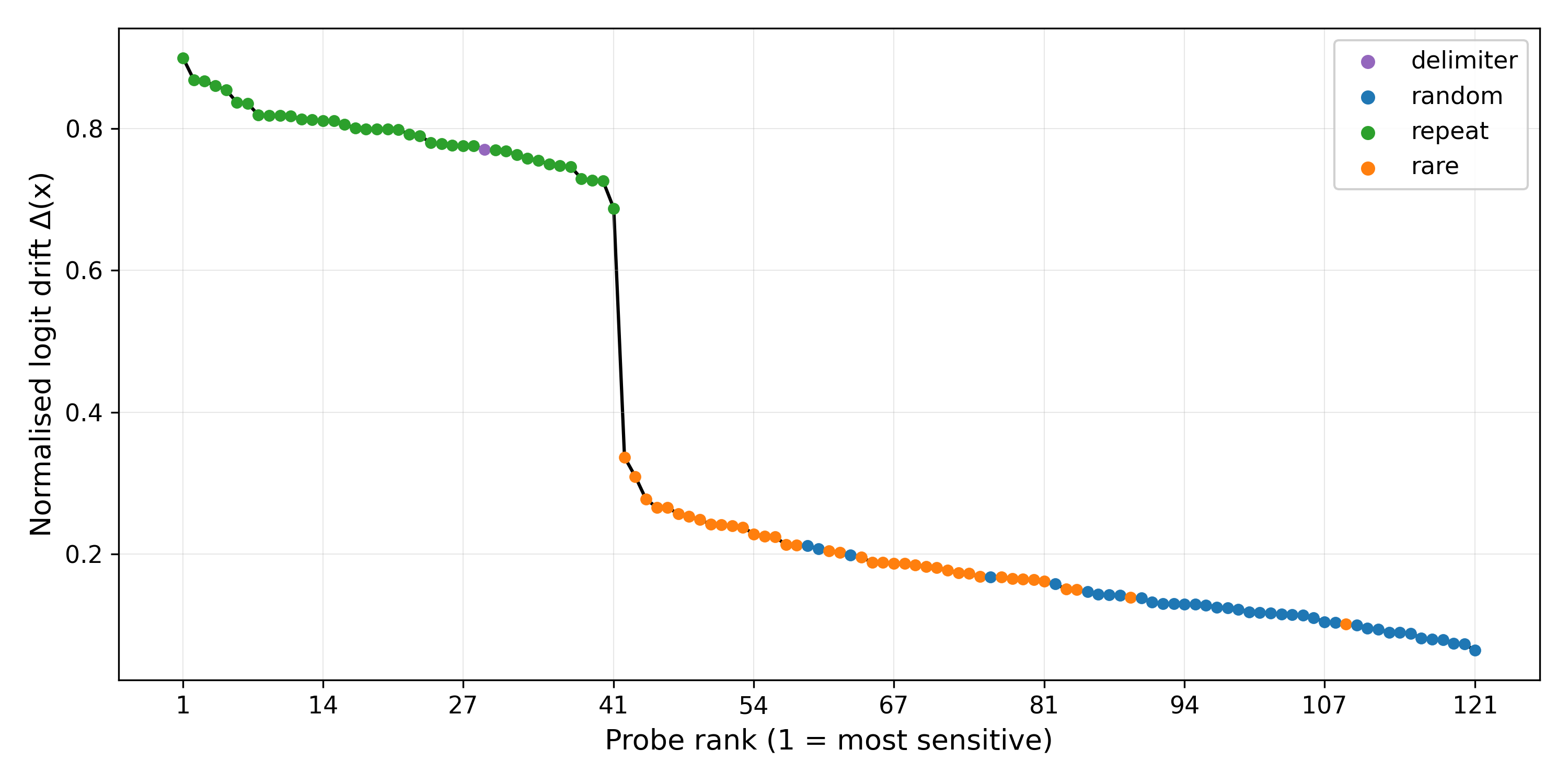}
  \caption{Ranked Distribution of Token-Based Probes by $\Delta(x)$}
  \label{fig:fam1k}
\end{figure}
\begin{figure}[tb]
  \centering
  \includegraphics[width=\linewidth]{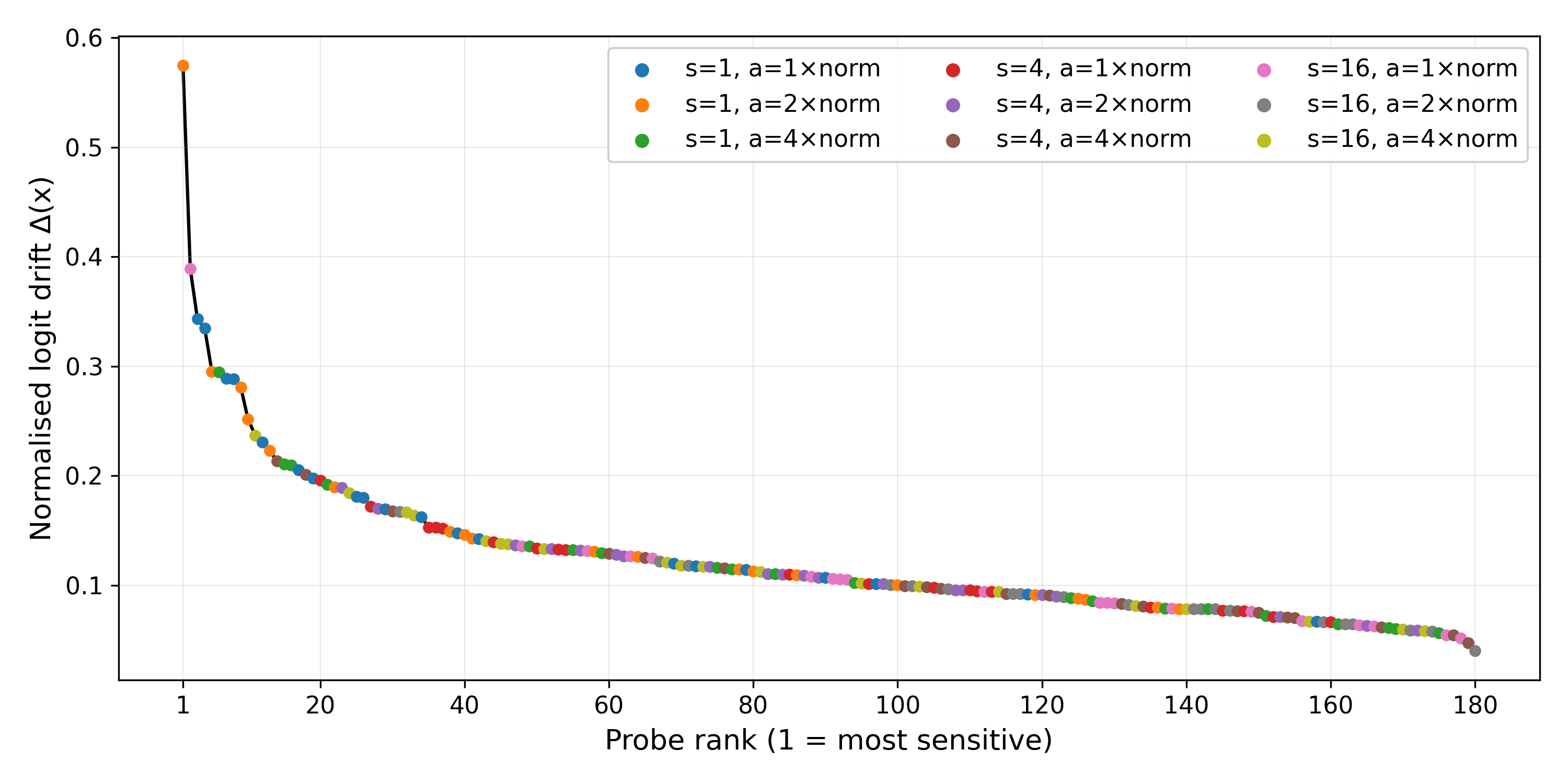}
  \caption{Ranked Distribution of Sparse Embedding Probes by $\Delta(x)$}
  \label{fig:fam2k}
\end{figure}
\begin{figure}[tb]
  \centering
  \includegraphics[width=\linewidth]{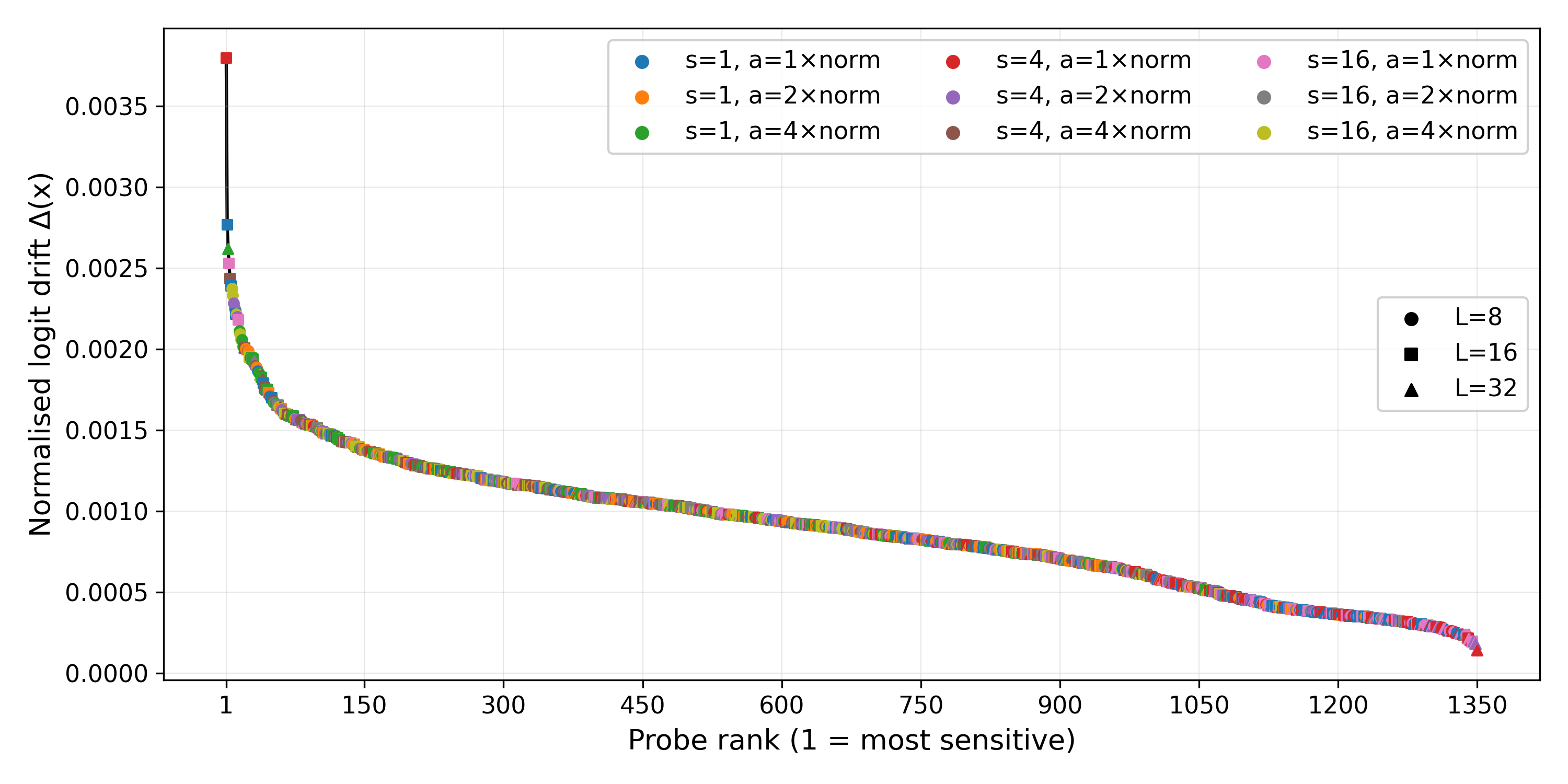}
  \caption{Ranked Distribution of MLP-Targeted Probes by $\Delta(x)$}
  \label{fig:fam3ak}
\end{figure}
\begin{figure}[tb]
  \centering
  \includegraphics[width=\linewidth]{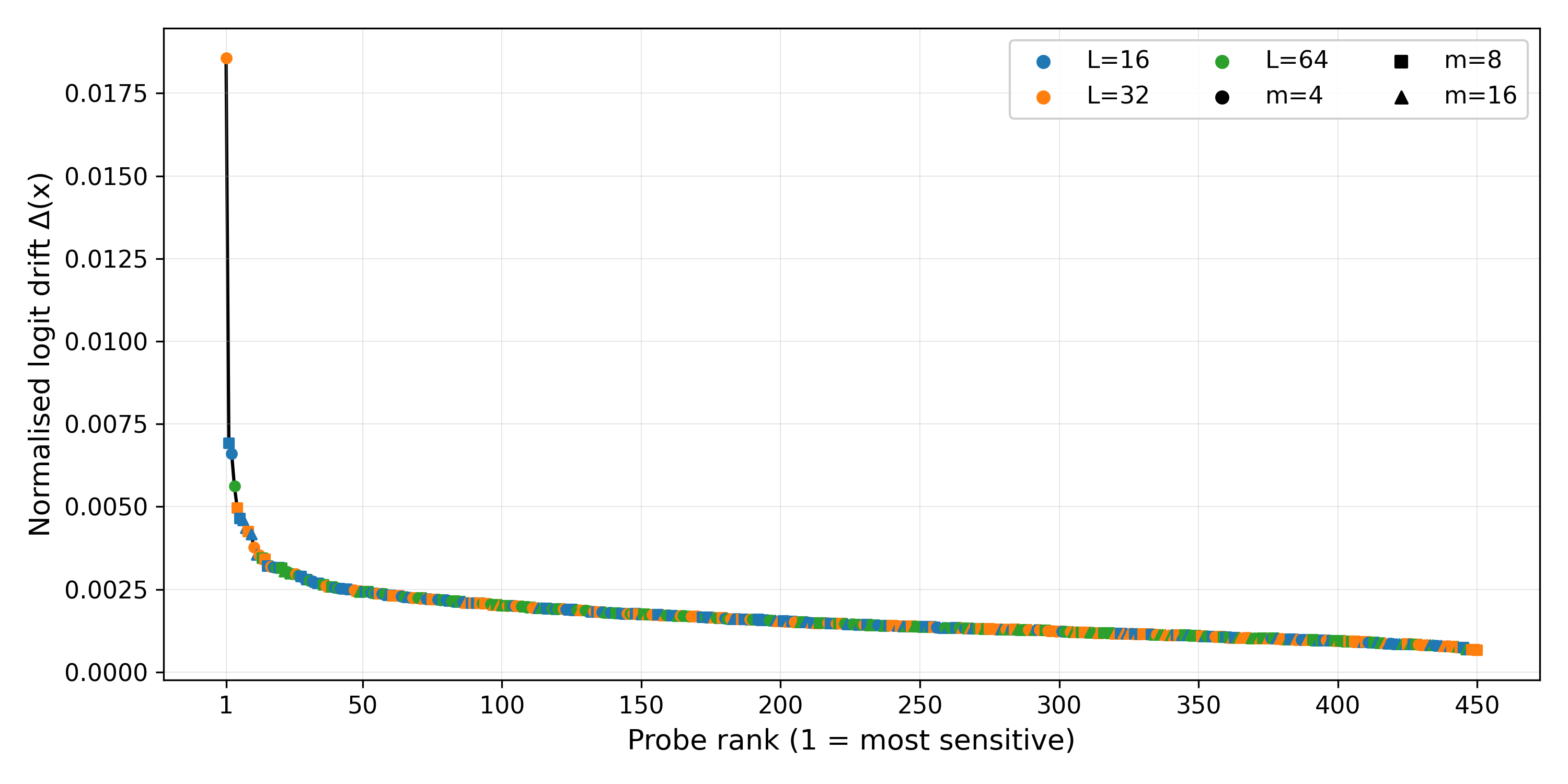}
  \caption{Ranked Distribution of Attention-Targeted Probes by $\Delta(x)$}
  \label{fig:fam3bk}
\end{figure}

Figures~\ref{fig:fam1k} through~\ref{fig:fam3bk} present the logit drift induced by each probe within its family, ranked in descending order such that rank 1 corresponds to the probe producing the greatest drift. Figure~\ref{fig:fam1k} shows the ranked distribution of token-based probes evaluated against the LoRA + Gaussian noise model. The Repeat probes induce the largest drift, with only the Delimiter probes producing comparable values, followed by a sharp falloff for the Rare and Random types. Each probe type occupies a clearly separated region of the distribution with little overlap, indicating that the drift behavior of each type is both stable and predictable.
 
In contrast, the ranked distribution of sparse embedding probes under the same LoRA + Gaussian noise model exhibits no clear trend, as shown in Figure~\ref{fig:fam2k}. Neither the number of nonzero dimensions $S$ nor the perturbation scale $A$ correlates with the observed drift. Moreover, the sparse embedding probes behave heterogeneously: the top 15 probes outperform the remaining 165 by a factor of three to six.
 
A similar pattern holds for MLP-targeted probes evaluated on the Late MLP model (Figure~\ref{fig:fam3ak}), where $S$, $A$, and the embedding length $L$ remain uncorrelated with logit drift. Likewise, the distribution of attention-targeted probes on the Early Attention model (Figure~\ref{fig:fam3bk}) reveals no clear relationship between sequence length $L$ or anchor spacing $M$ and the resulting drift. The MLP-targeted and attention-targeted probes additionally exhibit more pronounced outliers among their best-performing probes, with steep falloffs within the top five probes.
 
Taken together, these results highlight the stability of token-based probes relative to the other probe families. Although large gaps separate the token-based probe types, drift remains consistent within each type. The same cannot be said of the sparse embedding and stress-style probes, in which only a small fraction of the generated probes prove sensitive, followed by a sharp decline, and in which no parameter exerts a predictable effect on logit drift.

\subsection{Cross-Family Comparison under Each Perturbation Model}
\label{subsec:crossfam}

\begin{figure}[tb]
  \centering
  \includegraphics[width=\linewidth]{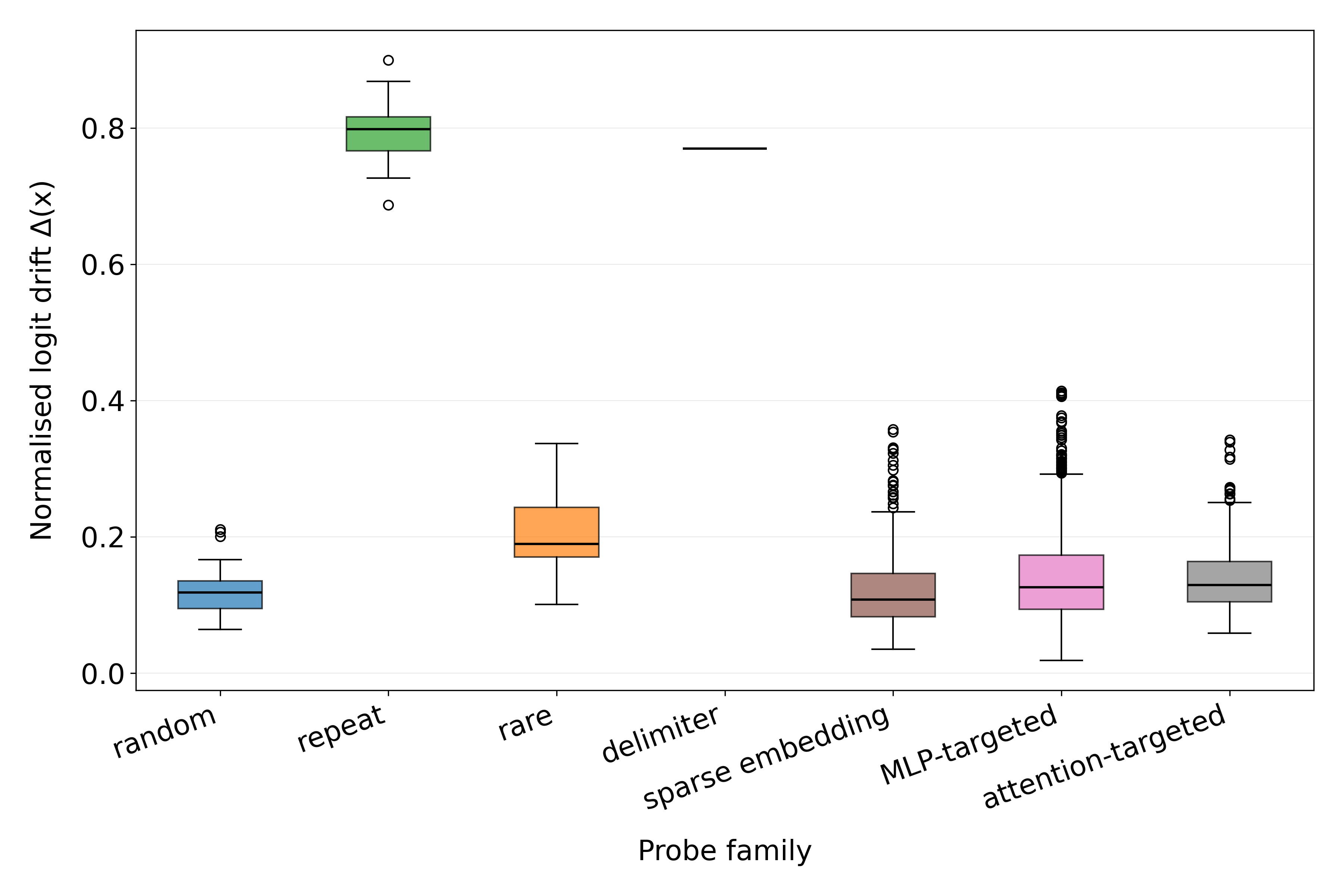}
  \caption{Comparison of $\Delta(x)$ Distributions by Probe Type on LoRA + Gaussian Llama 3.2 model}
  \label{fig:finalboxlora}
\end{figure}

\begin{figure}[tb]
  \centering
  \includegraphics[width=\linewidth]{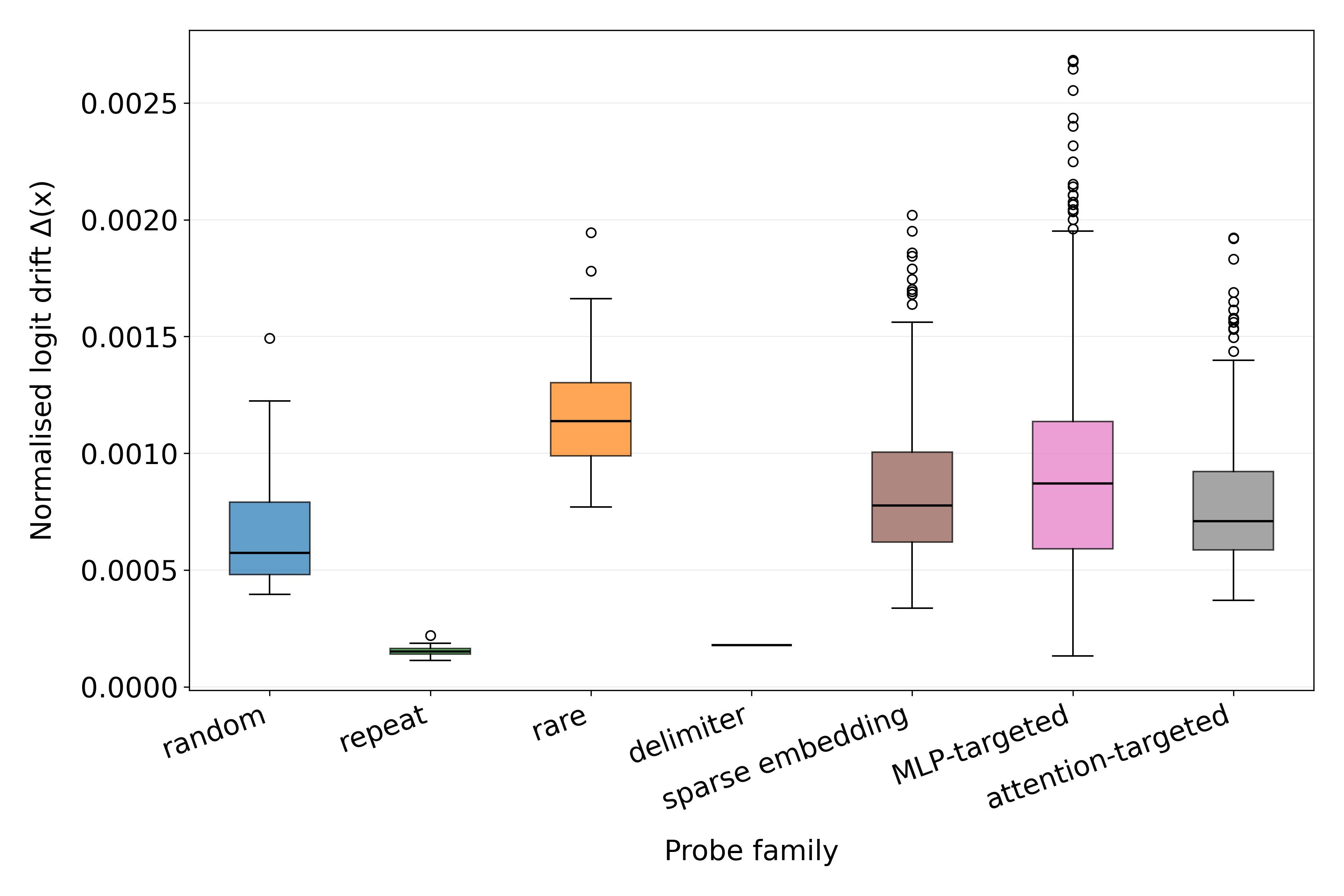}
  \caption{Comparison of $\Delta(x)$ Distributions by Probe Type on Late MLP Llama 3.2 model}
  \label{fig:finalboxmlp}
\end{figure}

\begin{figure}[tb]
  \centering
  \includegraphics[width=\linewidth]{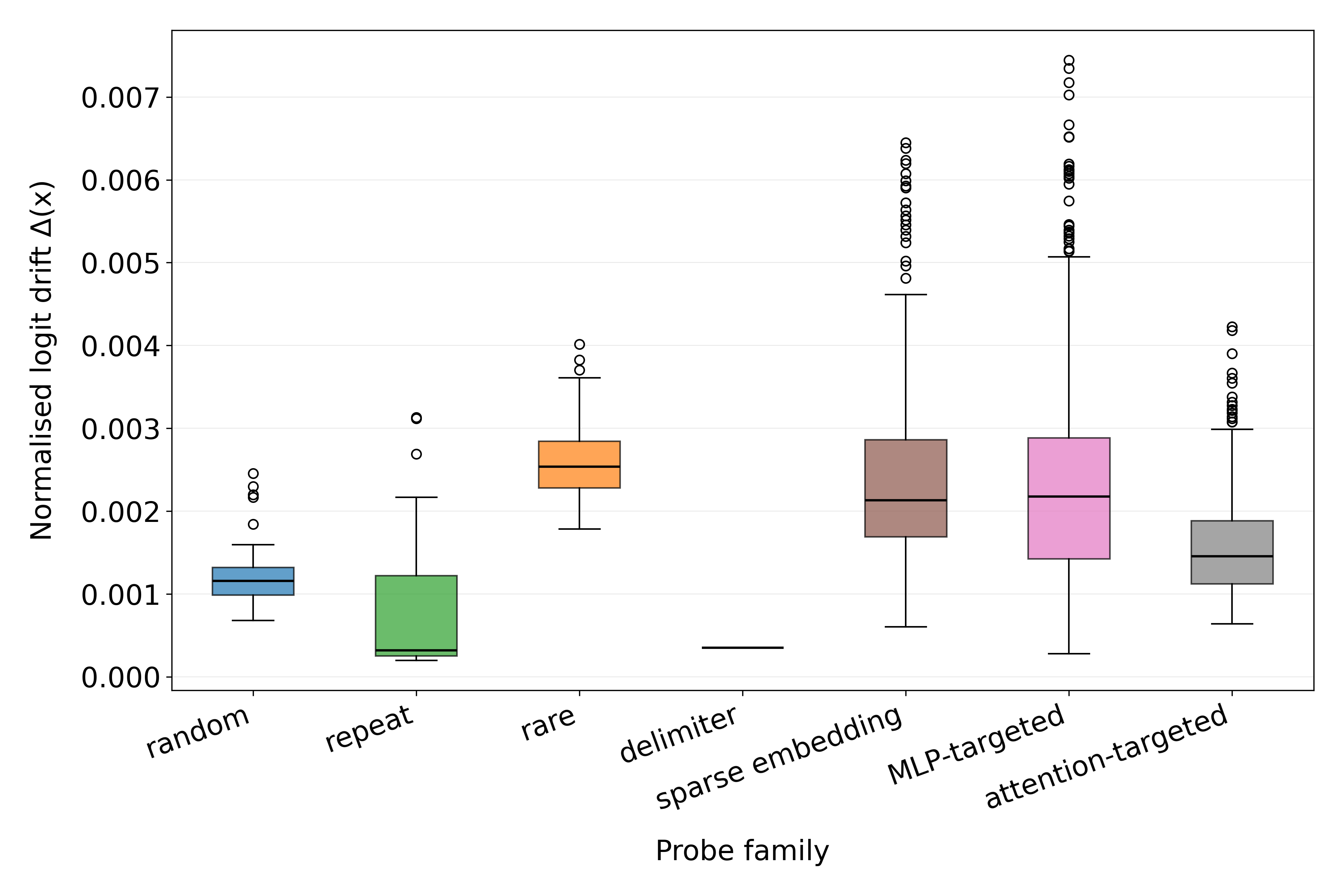}
  \caption{Comparison of $\Delta(x)$ Distributions by Probe Type on Early Attention Llama 3.2 model}
  \label{fig:finalboxatten}
\end{figure}

Figure~\ref{fig:finalboxlora} presents the cross-family $\Delta(x)$ comparison for the LoRA plus Gaussian perturbation model. The token-based probes exhibit the highest $\Delta(x)$, with the Repeat and Delimiter probes inducing drifts roughly four times as large as the remaining probes. The sparse embedding and MLP-targeted probes reveal a substantial number of outliers, consistent with the behavior described in Section~\ref{subsec:fam1}. Excluding these outliers, the MLP-targeted and attention-targeted probes perform comparably on average, while the sparse embedding probes perform the worst.

This ordering shifts under the localized perturbation models. In Figure~\ref{fig:finalboxmlp}, which reports the cross-family $\Delta(x)$ comparison for the Late MLP model, the token-based probes continue to perform well, yet Repeat and Delimiter become nearly ineffective, and Rare alone now induces drifts comparable to those of the sparse embedding and stress-style probes. Notably, the distributions for the sparse embedding and stress-style probes swell in this setting. Although the whiskers of the sparse embedding and MLP-targeted probes extend beyond those of Rare, Rare's mean drift remains considerably higher than that of all other families. The MLP-targeted probes follow the token-based probes in average drift but exhibit the highest outliers, with the sparse embedding and attention-targeted probes ranking third and fourth, respectively.

In the cross-family $\Delta(x)$ comparison for the Early Attention model (Figure~\ref{fig:finalboxatten}), a similar trend emerges. Here too, Rare outperforms all other families on average, though the sparse embedding and MLP-targeted probes again exhibit upper-percentile (80th–90th) results that exceed Rare. The MLP-targeted probes again outperform the sparse embedding probes, albeit by a much smaller margin, and the attention-targeted probes again yield the weakest results of any family. 

The localized perturbation experiments therefore do not support a pathway-specific advantage for the stress probes.
Under the late-MLP perturbation, the MLP-targeted family produces several strong outliers, but Rare token probes have the highest mean drift. 
Under the early-attention perturbation, the attention-targeted family is the weakest, while Rare again provides the highest mean drift. 
The stress constructions can therefore generate individual high-sensitivity candidates, particularly for the MLP perturbation, but they do not provide the expected family-level advantage. 
 
This result is stronger than the original pathway-specific hypothesis, i.e., simple token-space probes detect both broad and localized changes more consistently than probes explicitly motivated by the affected pathway. 
The probes within the family, however, vary considerably in effectiveness: Repeat and Delimiter, the strongest probes under the LoRA-plus-Gaussian model, become far less sensitive to the narrower changes induced by the Late MLP and Early Attention perturbations, whereas Rare, weak under LoRA-plus-Gaussian, matches or exceeds all other families on the remaining two models. 
As the token-based probes deliver consistently high $\Delta(x)$ values without the wide distribution of effectiveness exhibited by the sparse embedding and stress-style probes, we select them as the most effective probe family for Groth16 integration.

\subsection{Groth16 Performance}
\label{subsec:groth}

\begin{figure}[tb]
  \centering
  \includegraphics[width=\linewidth]{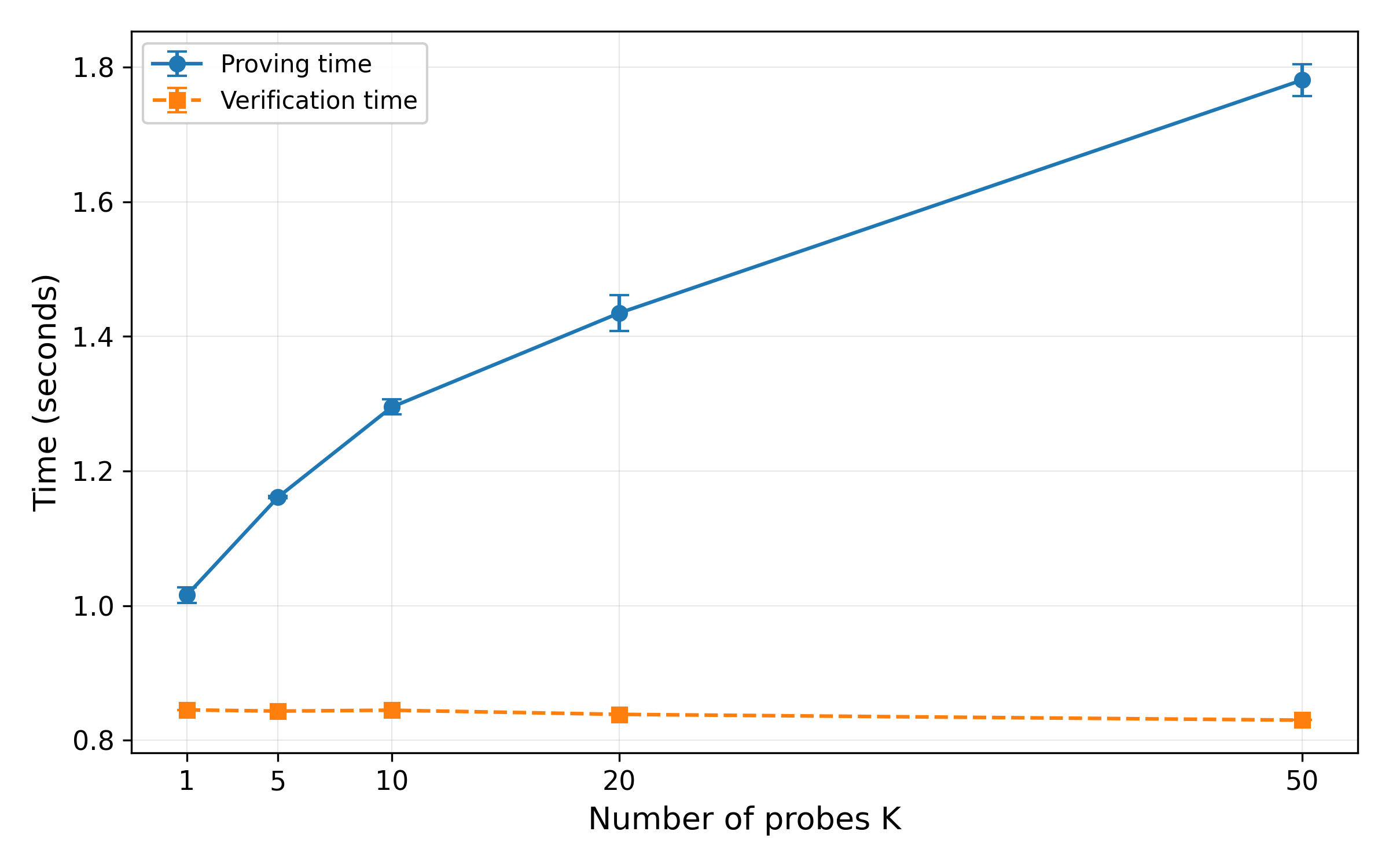}
  \caption{Proving and Verification Time vs Top-$K$ Probes Used}
  \label{fig:scalingproving}
\end{figure}

\begin{figure}[tb]
  \centering
  \includegraphics[width=\linewidth]{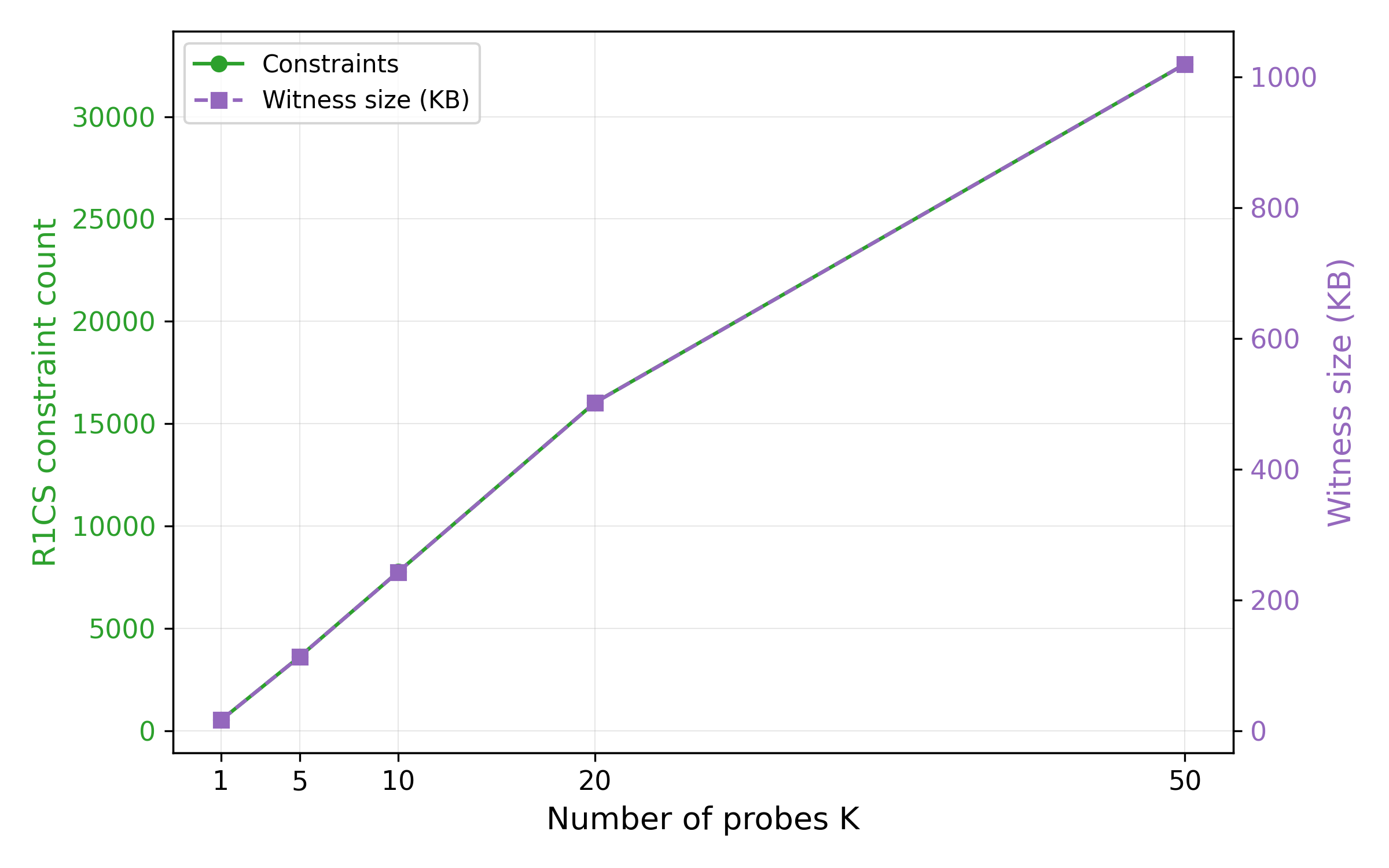}
  \caption{R1CS Constraints and Witness Size vs Top-$K$ Probes Used}
  \label{fig:scalingwitness}
\end{figure}

\begin{table*}[t]
\centering
\caption{Groth16 scaling results by number of probes $K$}
\label{tab:scaling_results}
\begin{tabular}{r r r r r r r}
\hline
$K$ & Constraints & Prove (s) & Verify (s) & Witness (KB) & Proof (bytes) & Audit (s) \\
\hline
1  &   517   & $1.016 \pm 0.01$ & $0.845 \pm 0.00$ &   16.3  & 805 & 1.891 \\
5  &  3619   & $1.161 \pm 0.00$ & $0.843 \pm 0.00$ &  113.4  & 804 & 2.157 \\
10 &  7755   & $1.295 \pm 0.01$ & $0.845 \pm 0.00$ &  242.8  & 804 & 2.445 \\
20 & 16027   & $1.435 \pm 0.03$ & $0.838 \pm 0.00$ &  501.6  & 804 & 2.883 \\
50 & 32571   & $1.781 \pm 0.02$ & $0.830 \pm 0.00$ & 1019.5  & 805 & 4.136 \\
\hline
\end{tabular}
\end{table*}

Having selected the token-based probes under the primary Llama LoRA-plus-Gaussian calibration condition, we rank them by $\Delta(x)$ and retain the top $K$ probes for the Groth16 cost evaluation. 
The circuit measurements study the cost of committing to a fixed probe set and do not assume that this particular set is optimal for every model or perturbation.

To identify the optimal $K$, we evaluated the cost of proving and verifying the hidden probe audit. The circuit design, a Merkle tree of Poseidon~\cite{poseidon} hashes, varies only in the number of inputs to the circuit, and the probe set remains fixed for each experiment. 
In our first experiment, we ran each probe count through the circuit five times to account for variance in proving time. Figure~\ref{fig:scalingproving} shows that proving time scales sub-linearly with $K$, with proving time less than doubling between $K=1$ and $K=50$. As expected from Groth16, verification time remains stable at roughly 0.84 seconds regardless of $K$. The bulk of the audit cost is therefore borne by the prover rather than the verifier, regardless of $K$. Figure~\ref{fig:scalingwitness} reports a second experiment, in which we measured the R1CS constraint count and witness size against $K$. Both grow sharply with $K$: at $K=50$, the witness size is 65 times that of $K=1$. However, $K=50$ requires only a witness of 1~MB, so the size of the witness does not constitute a substantial bottleneck for this system.

Table~\ref{tab:scaling_results} provides a breakdown of all data presented in these figures, together with the audit time, defined as the combined time for inference, witness preparation, proving, and verification; this constitutes the full time required to prove the system and matches the cost incurred during commitment. At low $K$, audit time is dominated by proving and verification, but inference time grows rapidly as $K$ increases.

Because token-based probes produce predictably high drift, a large $K$ is not required; between 5 and 20 probes suffice to minimize audit time while preserving accuracy. This yields a total audit time of between 2.157 and 2.883 seconds. 
Our evaluation isolates the cryptographic cost of increasing \(K\). 
The range \(K=5\) to \(20\) should be interpreted as a practical cost range. 
The appropriate \(K\) for detection can be easily calibrated separately for the model and deployment setting. 

\subsection{Cross-Hardware Validation}
\label{subsec:hardware}

The preceding results were collected on a single GPU, the NVIDIA RTX A4000. Because our system is intended to run across many different hardware configurations, drift signatures that hold on one device but not another would undermine its usefulness. To this end, we repeated the probe evaluation on five NVIDIA GPUs — the A100, RTX A6000, L40S, H100, and H200 — covering workstation, inference-optimized, and datacenter-class hardware.

\begin{figure}[tb]
  \centering
  \includegraphics[width=\linewidth]{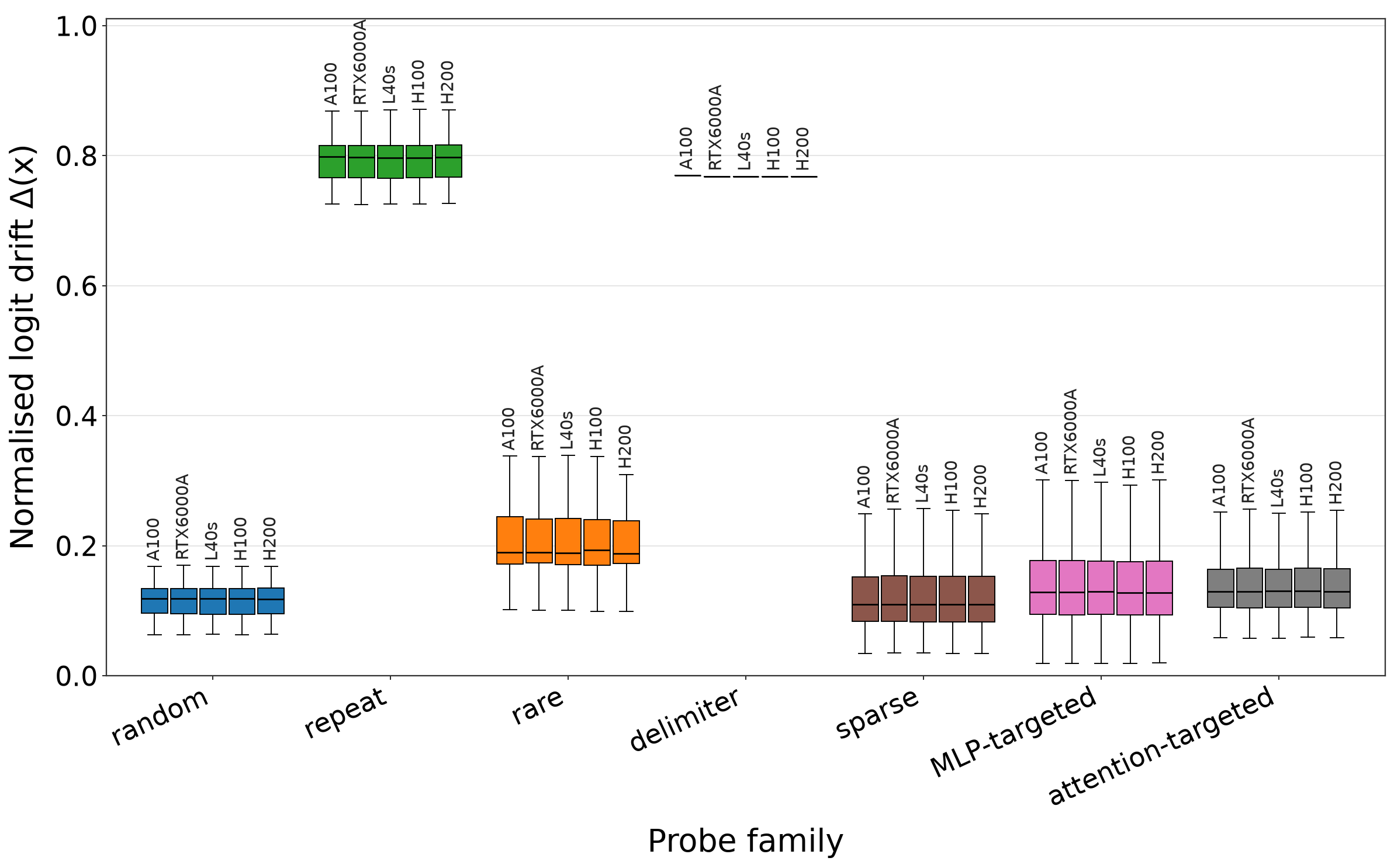}
  \caption{Comparison of $\Delta(x)$ Distributions by Probe Type on LoRA + Gaussian Llama~3.2 model across Multiple GPUs}
  \label{fig:loragpus}
\end{figure}

\begin{figure}[tb]
  \centering
  \includegraphics[width=\linewidth]{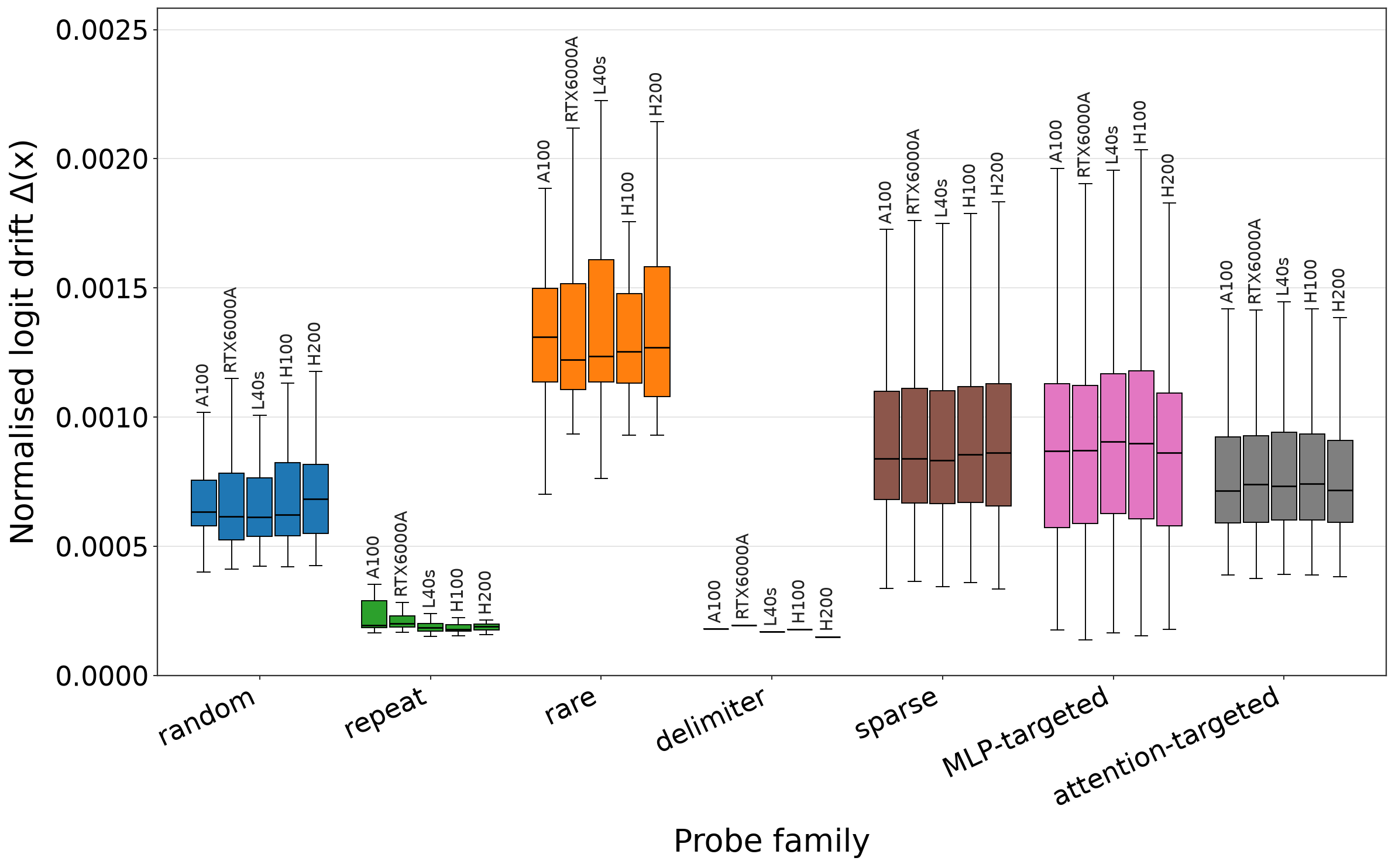}
  \caption{Comparison of $\Delta(x)$ Distributions by Probe Type on Late MLP Llama 3.2 model across Multiple GPUs}
  \label{fig:lategpus}
\end{figure}

\begin{figure}[tb]
  \centering
  \includegraphics[width=\linewidth]{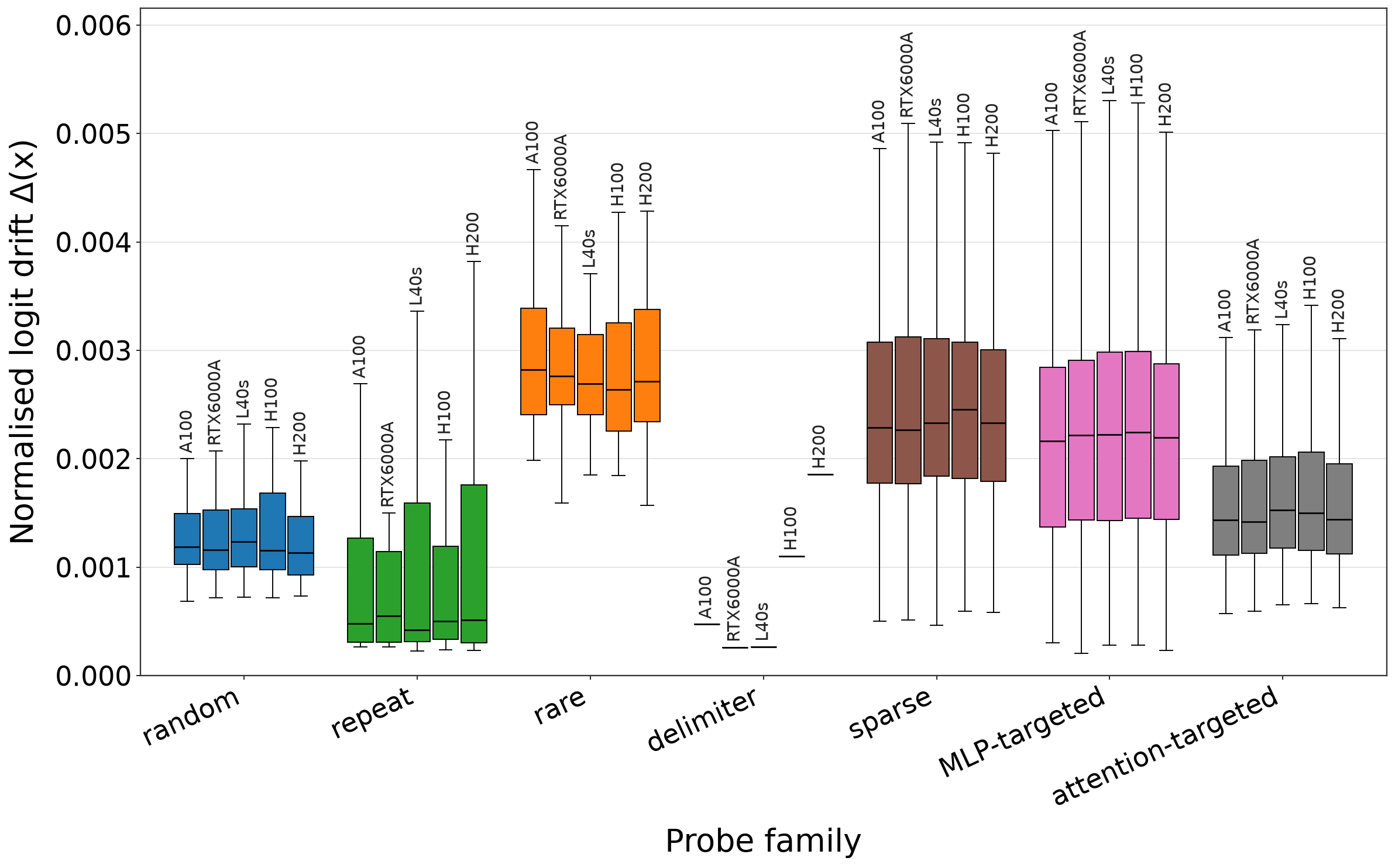}
  \caption{Comparison of $\Delta(x)$ Distributions by Probe Type on Early Attention Llama 3.2 model across Multiple GPUs}
  \label{fig:earlygpus}
\end{figure}

The cross-family $\Delta(x)$ comparison for the LoRA plus Gaussian Noise model between different GPUs (Figure~\ref{fig:loragpus}) shows only minute differences between GPU runs.  Small variations in medians and upper/lower quartiles are present in each probe family, but the ranking of families still remains the same, and each GPU presents a signature for the model that is very similar.

Figure~\ref{fig:lategpus} shows a less clean picture for the performance on the Late MLP model between GPUs.  While the overall rankings of the probe families do not change on any of the models, there are substantial differences in how each GPU processed a given probe.  Within the Repeat family, for example, the upper quartile probes performed considerably better on the A100 than on any other GPU tested.

In Figure~\ref{fig:earlygpus}, the GPUs again differ in probe performance when running the Early Attention model. Here, several rankings change depending on the GPU.  The Rare family still has the median best-performing probes across all GPUs, but on the H200 the upper quartile of Repeat probes exceeds that of the Random probes — a relationship that reverses on the H100.

We expect these differences seen in Figures~\ref{fig:lategpus} and~\ref{fig:earlygpus} to be caused by different architectures, as the LoRA plus Gaussian Noise model, in which all layers are affected, performs the same across all GPUs.  
In fact, across these experiments, we held the baseline settings fixed. 
The GPU device was the only experimental factor that changed. 
Hence, a deployment may require a device-specific baseline. 

\subsection{Cross-Model Validation of Probe Selection}
\label{subsec:Qwen}
\begin{figure}[tb]
  \centering
  \includegraphics[width=\linewidth]{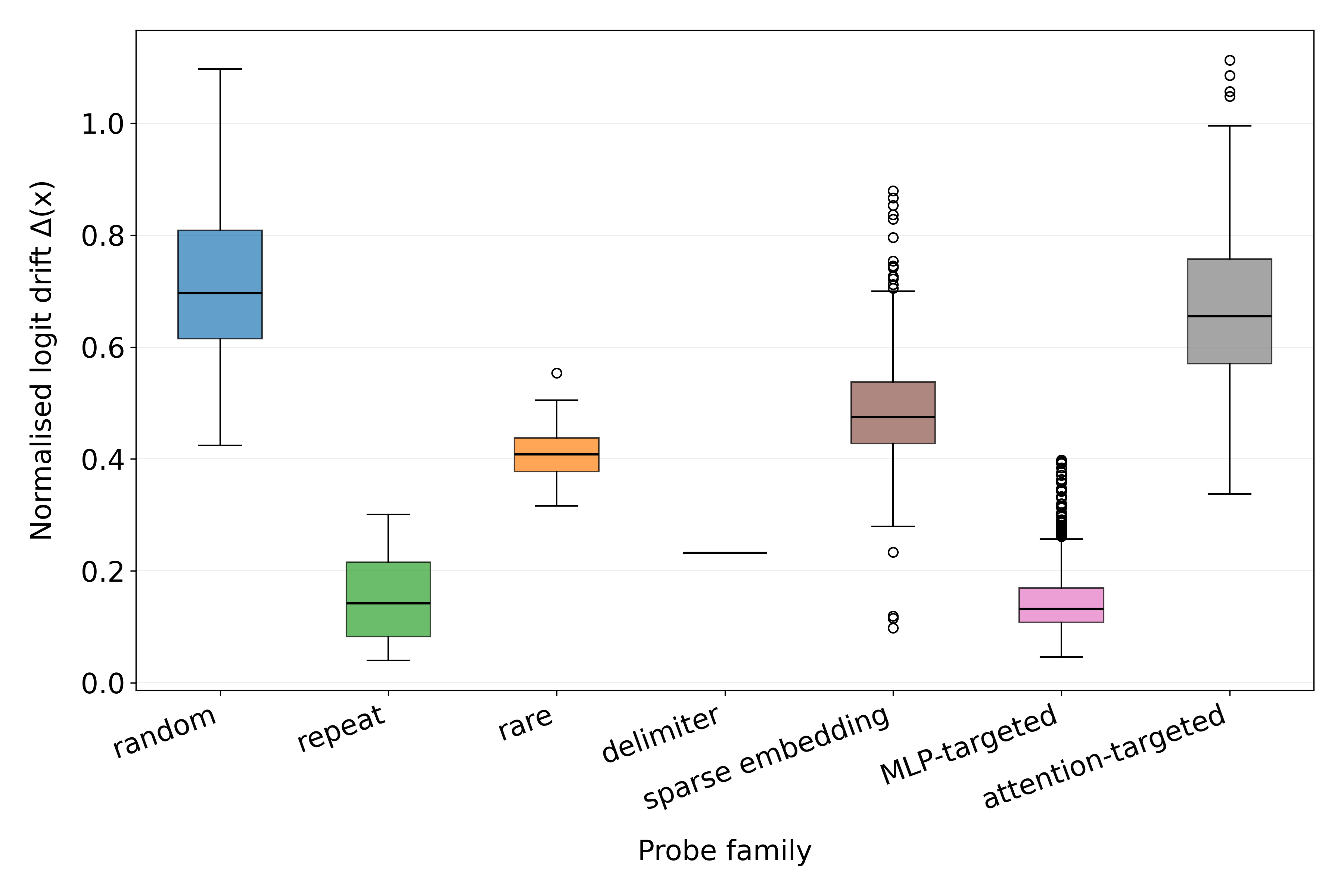}
  \caption{Comparison of $\Delta(x)$ Distributions by Probe Type on LoRA + Gaussian Qwen 2.5 model}
  \label{fig:finalqwenlora}
\end{figure}

\begin{figure}[tb]
  \centering
  \includegraphics[width=\linewidth]{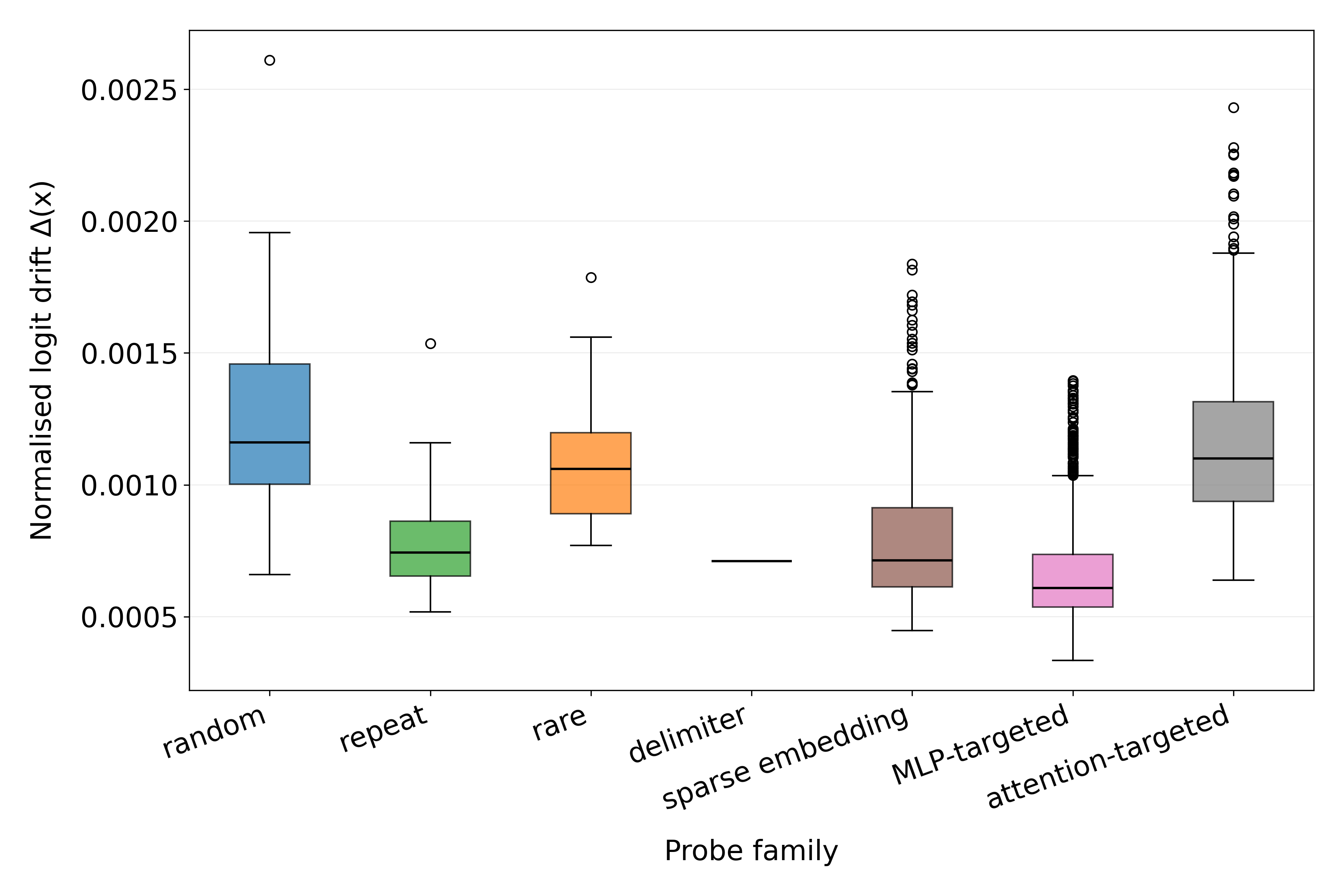}
  \caption{Comparison of $\Delta(x)$ Distributions by Probe Type on Late MLP Qwen 2.5 model}
  \label{fig:finalqwenmlp}
\end{figure}

\begin{figure}[tb]
  \centering
  \includegraphics[width=\linewidth]{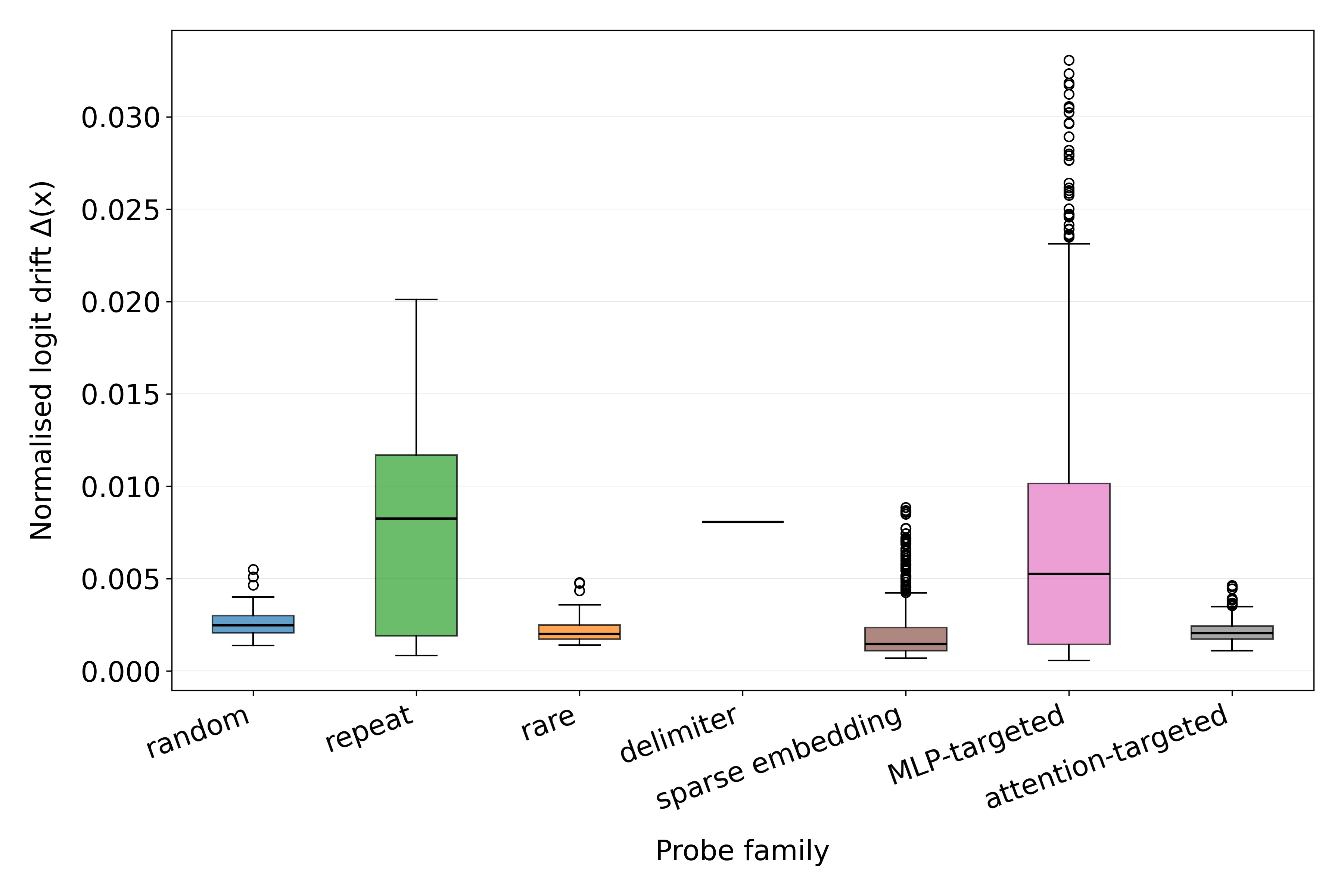}
  \caption{Comparison of $\Delta(x)$ Distributions by Probe Type on Early Attention Qwen 2.5 model}
  \label{fig:finalqwenatten}
\end{figure}

To determine whether the strength of the token-based probes reflects the model architecture or a more general property, we repeated the experiment on Qwen2.5-1.5B-Instruct, chosen for its dissimilarity to Llama-3.2-1B, on the original A4000 GPU. Figures~\ref{fig:finalqwenlora},~\ref{fig:finalqwenmlp}, and~\ref{fig:finalqwenatten} report the response of each probe type to the three perturbation models.

Under the LoRA plus Gaussian Noise perturbation (Figure~\ref{fig:finalqwenlora}), the strongest probes remain among the token-based probes, though the leader shifts: Random now produces the highest mean logit drift, with the attention-targeted probes following closely. The wide separation between Repeat, Delimiter, and the remaining token probes seen on Llama is absent, and the margin between families narrows.

The intra-family ordering diverges further under the localized perturbations. In Figure~\ref{fig:finalqwenmlp}, perturbing the Qwen MLP layer causes Rare,  the strongest token probe for MLP changes on Llama, to fall behind Random, which produces the highest mean drift. 
The attention-targeted probes are nearly indistinguishable from Random: several of their outliers exceed it, and only a single Random outlier separates the two. Under the Early Attention perturbation (Figure~\ref{fig:finalqwenatten}), Repeat produces the highest mean drift, though the MLP-targeted probes trail closely and reach further in their upper outliers. 
Accordingly, the MLP-targeted and attention-targeted labels should be interpreted as probe-generation hypotheses rather than as empirically verified pathway-specific detectors. 

To summarize, the experiments on Qwen provide a clear overview of how to select the most appropriate probe family. 
They highlight the fact that the most sensitive probe must be identified by searching a heterogeneous synthetic candidate pool for each model. 
Robust probe selection is therefore a per-model calibration step, meaning that the appropriate probe must be re-established for each model, not carried over from another architecture.

\section{Conclusion}
We presented a hidden-probe attestation framework that verifies model integrity against a baseline commitment without revealing the probes themselves, binding a set of drift-sensitive probes to a Groth16 proof over a Poseidon Merkle-tree commitment. 
Our evaluation identified token-based probes as the most broadly effective family, and the Qwen 2.5 experiments confirmed token-space search continues to identify the highest-mean probe type across the two architectures, although the effective construction changes by model and perturbation.  
The behavior was further reproduced across GPUs, and the full audit completes in roughly 2.4 seconds end-to-end at $K$=10, with verification fixed near 0.84 seconds. 
These results suggest that hidden-probe attestation can serve as a practical mechanism for repeated, low-overhead verification of deployed models. 
Several limitations remain. Our evaluation considered two small models and a fixed set of perturbation types, and both the leading probe and the drift baseline must be re-established per model and per device rather than reused across them. 
We also do not claim resistance against an adversary who adapts to the probe-selection procedure itself. 
Extending the framework to larger and more diverse models, characterizing its detection guarantees under a formal threat model, and reducing the per-deployment calibration cost remain important directions for future work.


\section*{Acknowledgments}
This effort was partially sponsored by NSF Grants CNS-2541809, Longview Philanthropy, Hardware-Enabled Mechanisms (HEMs), as well as Survival \& Flourishing Fund (SFF-2024) Mechanisms for Flexible
Hardware-Enabled Guarantees (flexHEGs). 

\small
\bibliographystyle{ACM-Reference-Format}
\bibliography{references}

\end{document}